\documentclass{article}
\usepackage[ansinew]{inputenc}
\usepackage{latexsym}
\newcommand\bearst{\begin{eqnarray*}}
\newcommand\eearst{\end{eqnarray*}}

\newcommand{\slpe}{\raise.15ex\hbox{$/$}\kern-.57em\hbox{$p$}}
\newcommand{\slpartial}{\raise.15ex\hbox{$/$}\kern-.57em\hbox{$\partial$}}
\newcommand{\slp}{\raise.15ex\hbox{$/$}\kern-.57em\hbox{$p$}}
\newcommand{\slq}{\raise.15ex\hbox{$/$}\kern-.57em\hbox{$q$}}
\newcommand{\slk}{\raise.15ex\hbox{$/$}\kern-.57em\hbox{$k$}}
\newcommand{\sla}{\raise.15ex\hbox{$/$}\kern-.57em\hbox{$a$}}
\newcommand{\slA}{\raise.15ex\hbox{$/$}\kern-.57em\hbox{$A$}}
\newcommand{\slB}{\raise.15ex\hbox{$/$}\kern-.57em\hbox{$B$}}
\newcommand{\slD}{\raise.15ex\hbox{$/$}\kern-.57em\hbox{$D$}}
\newcommand{\slb}{\raise.15ex\hbox{$/$}\kern-.57em\hbox{$b$}}
\newcommand{\slc}{\raise.15ex\hbox{$/$}\kern-.57em\hbox{$c$}}
\newcommand{\sld}{\raise.15ex\hbox{$/$}\kern-.57em\hbox{$d$}}
\newcommand{\slW}{\raise.15ex\hbox{$/$}\kern-.57em\hbox{$W$}}
\newcommand{\slP}{\raise.15ex\hbox{$/$}\kern-.57em\hbox{$P$}}

\newcommand{\be}{\begin{equation}}
\newcommand{\ee}{\end{equation}}
\newcommand{\bear}{\begin{eqnarray}}
\newcommand{\ear}{\end{eqnarray}}
\newcommand{\ba}{\begin{eqnarray*}}
\newcommand{\ea}{\end{eqnarray*}}
\renewcommand{\theequation}{\arabic{section}.\arabic{equation}}

\newcommand{\dt}{\mbox{\boldmath$:$}}

\newcommand{\no}{\noindent}

\newcommand{\z}{\mathit{z}}

\begin{document}

\title{Two-Dimensional Thermofield Bosonization II: Massive Fermions}
\author{R. L. P. G. Amaral$^\ast$, L. V. Belvedere$^\ast$ and K. D. Rothe$^{\ast\ast}$\\
\small{$\ast$ Instituto de F\'{\i}sica}\\
\small{Universidade Federal Fluminense}\\
\small{Av. Litor\^anea S/N, Boa Viagem, Niter\'oi, CEP. 24210-340}\\
\small{Rio de Janeiro - Brasil}\\
\small{$\ast \ast$ Institut f\"ur Theoretische Physik}\\
\small{Universit\"at Heidelberg}\\
\small{Philosophenweg 16, D-69120 Heidelberg}\\
\small{Germany}\\
}

\date{\today}

\maketitle

\begin{abstract}
We consider the perturbative computation of the N-point function of chiral densities of massive free fermions at finite temperature within the thermofield dynamics approach. The infinite series in the mass parameter for the N-point functions are computed in the fermionic formulation and compared with the corresponding perturbative series in the interaction parameter in the bosonized thermofield formulation. Thereby we establish in thermofield dynamics the formal equivalence of the massive free fermion theory with the sine-Gordon thermofield model for a particular value of the sine-Gordon parameter. We extend the thermofield bosonization to include the massive Thirring model.  
\end{abstract}

\section{Introduction}

The bosonization of fermions has proven in the past to be a very useful technique for solving quantum field theoretic models in 1+1 dimensions \cite{AAR}. In a recent paper \cite{ABR} we have considered the operator formulation of the bosonic representation of massless free fermions at finite temperature (thermofield bosonization) with the thermofield dynamics formalism \cite{TFD,U,Das,Haag,Mats1,O,Mats}. The well known two-dimensional Fermion-Boson correspondences at zero temperature are shown to hold also at finite temperature. In Ref. \cite{ABR} we have also used the thermofield bosonization for obtaining the real time fermion N-point functions of the massless Thirring model at finite temperature.

The equivalence of the massive Thirring model and the sine-Gordon theory at finite temperature in the imaginary time formalism has been the subject of a number of authors \cite{Delep,Imag}. The discussions have been carried out predominantly from the functional point of view.

The main objective of the present work is to extend the thermofield bosonization approach presented in Ref. \cite{ABR} to the case of massive fermions. To this end we use strictly fermionic techniques on the one hand, and bosonization
techniques on the other, in order to demonstrate within the Thermofield Dynamics approach the equivalence of the theory of massive free
fermions at finite temperature with the sine-Gordon thermofield theory. The demonstration will be done for the case of the
N-point functions of chiral densities, by working in the interaction picture of the respective formulations, with the mass term playing the role of the interaction Hamiltonian. 
The selection rule to be imposed in both formulations (fermionic and bosonic) in order to prove the equivalence emerges here in an interesting way. As a byproduct the role of the ``tilde'' fields in thermofield dynamics
and the lower branch at $t \,= \,-\, i \,\beta/2$ in the real time formalism \cite{Mats1} emerges naturally in this
calculation. 

The intimate relationship between the thermofield dynamics formalism and the algebraic formulation due to Haag-Hugenholtz-Winnink (HHW) of statistical mechanics  has been established in Ref. \cite{O}. In this reference the relevance of tilde objects to the modular conjugation appearing in the algebraic formulation of statistical mechanics in the HHW formalism based upon the KMS condition is clarified. This formulation of thermofield dynamics in a way consistent with the HHW formalism enable to extend it to gauge theories and becomes crucial in the treatment of the Faddeev-Popov ghosts \cite{O}. To this end, a ``new'' version for the thermofield dynamics approach for fermions is presented in \cite{O}. In this paper we shall follow this revised version for the thermofield dynamics approach for fermions since it corrects a mistake in our previous work \cite{ABR} (this is discussed in Appendix) and as we shall see, becomes fundamental for the success of the thermofield bosonization scheme.

The paper is organized as follows: We begin in Section 2 by considering the thermofield dynamics approach to the massive free fermion theory. We use the generalization of the perturbation theory to finite temperature \cite{Mats} in order to compute in thermofield dynamics the N-point function of chiral densities of massive free fermions as a power series in the mass $M$ of the fermion, with an explicit expression for the expansion coefficients as a ratio of temperature dependent polynomials. In Section 3 we repeat this calculation for the bosonized formulation and extend to include the massive Thirring model. The equivalence of the two formulations is thereby established, upon taking suitable account of the selection rule emerging in this calculation. In Section 4 we discuss the physical meaning of the selection rule.  We conclude in Section 5 with some comments. In Appendix  we discuss some modifications of the thermofield dynamics formulation for fermions of Ref. \cite{O} due to Ojima, which shall play an important role in order to obtain the correct bosonized expression for the Fermi thermofields. This streamlines the presentation of Ref. \cite{ABR}.

\section{Massive Free Fermions}

In thermofield dynamics the construction of a field theory at finite temperature requires doubling the numbers of fields degrees of freedom by introducing ``tilde'' operators corresponding to each of the operators describing the system considered \cite{TFD,U,Das,Haag,Mats1,O,Mats}. To this end, let us consider the total Lagrangian of the two-dimensional massive free Fermi field corresponding to the fermion doublet $\pmatrix{\psi \cr - i {\widetilde\psi}^\dagger}$ 

\be
\widehat{\cal L} = {\cal L} - \widetilde{\cal L}\,,
\ee

\no where ${\cal L}$ is the usual Lagrangian of a massive free fermion, and $\widetilde{\cal L}$ is the corresponding Lagrangian in terms of the field $\widetilde\psi$ and obtained from ${\cal L}$ by the tilde conjugation defined by $\widetilde{(c \psi)}\,=\,c^\ast \widetilde \psi$. Since we shall consider perturbation theory around the massless theory, we choose to write $\widehat{\cal L}$ in the form

\be
\widehat{\cal L} = \widehat{\cal L}_0 + \widehat{\cal L}_I
\ee

\no where \footnote{The conventions used 
are: 

$$\gamma^0 = \pmatrix{0 & 1 \cr 1 & 0}\,,\gamma^1 
= \pmatrix{0 & 1 \cr - 1 & 0}\,,\gamma^5 = \gamma^0 \gamma^1\,\,,\,\,
\epsilon_{0 1} = 1\,,\,g^{00} = 1\,,\,x^\pm = x^0 \pm x^1\,,\,\partial_\pm 
= \partial_0 \pm \partial_1\,.
$$ 

\no For massless scalar field $\varphi (x) = \phi (x^+) + \phi (x^-)$, and for the pseudo-scalar field $\phi (x) = \phi (x^+) - \phi (x^-)$.}

\be
\widehat{\cal L}_0 = \overline{\psi} i \gamma^\mu\partial_\mu \psi + \overline{\widetilde{\psi}} i \gamma^\mu\partial_\mu {\widetilde\psi}\,,
\ee

\no and 

\be\label{LI}
\widehat{\cal L}_I =\,-\, M(\overline{\psi}\psi - \overline{\widetilde\psi} \widetilde\psi)\,.
\ee

At finite temperature and within the thermofield approach, the perturbative computations can be performed using the generalization of the Gell-Mann-Low formula for $T \neq 0$ \cite{Mats}. The vacuum expectation value of time-ordered products of Heisenberg operators $\Phi$ in the physical vacuum $\vert 0 (\beta ) \rangle $ at finite temperature \cite{Mats1,O,Mats} is given by

\begin{equation}\label{general}
\langle 0(\beta) \vert \, T\,\prod_k{ \Phi}_k (x_k) \vert 0 (\beta)\rangle = \frac{\displaystyle \langle 0,\beta|T\prod_k\phi_k (x_k)\exp(i\int d\,^2 z \hat {\cal L}_I (z))|0,\beta\rangle}{\displaystyle\langle 0,\beta|T\exp(i\int d\,^2 z \hat {\cal L}_I (z))|0,\beta\rangle},
\end{equation}

\no where the right hand side is computed in the interaction picture.  The thermal interaction picture vacuum  is defined by

\be
\vert 0, \beta \rangle = U_F [\theta_F]\,\vert \widetilde 0, 0\rangle\,,
\ee

\no where $\vert \widetilde 0, 0 \rangle$ is the interaction picture vacuum at zero temperature, and
$U_F [\theta_F]$ is the unitary operator (We shall adopt the ``revised'' thermofield dynamics approach for fermions as formulated in Ref. \cite{O}) 

\be\label{Uf}
U_F [\theta_F ]\,=\,e^{\,-\,i\,\int_{-\infty}^\infty \,d\,p\,\theta_F (\vert p^1\vert  ,\beta )\Big ( \widetilde b (p^1)\,
b (p^1)\,+\,b^\dagger (p^1)\,\widetilde{b}^\dagger (p^1)\,+\, 
\widetilde d (p^1)\,
d (p)\,+\,d^\dagger (p^1)\,\widetilde{d}^\dagger (p^1)\,\Big )}\,,
\ee

\no where  the  Bogoliubov parameter $\theta_F (p , \beta )$  is implicitly defined by ($p = \vert p^1 \vert$),

\be
\cos \theta_F (p ; \beta )\,=\,
\frac{1}{\sqrt{1 + e^{\,-\,\beta\,p}}}\,,
\ee

\no and 

\be
\sin \theta_F (p ; \beta )\,=\,
\frac{e^{\,-\,\beta\,p /2}}{\sqrt{1 + e^{\,-\,\beta\,p}}}\,,
\ee

\no with the Fermi-Dirac statistical  weight given by,

\be
N_F (p; \beta )\, = \,\sin^2 \theta_F (p ; \beta )\,=\,\frac{1}{e^{\,\beta p} + 1}\,.
\ee

\no We can pass the time independent unitary operator (\ref{Uf}) through the time ordering operation, and rewrite the Gell-Mann-Low formula in terms of the zero temperature vacuum $\vert \widetilde 0 , 0 \rangle$ and the transformed annihilation operators are given by \cite{O}

\be
b (p ; \beta ) = b(p) \cos \theta_F (p ; \beta ) \,+\,i\,\widetilde b^\dagger (p) \sin \theta_F (p ; \beta )\,,
\ee

\be
\widetilde b (p ; \beta ) = \widetilde b(p) \cos \theta_F (p ; \beta ) \,-\,i\, b^\dagger (p) \sin \theta_F (p ; \beta )\,,
\ee

\no and their adjoints, with similar expressions for $d$ and $\widetilde d$. For free massless fermions in $1 + 1$ dimensions the spinor components $\psi_1 (\psi_2)$ are left (right) moving fields:

\be
\psi(x) = \left(
\begin{array}{c}
\psi_1(x^+)\\
\psi_2(x^-)\\
\end{array}
\right)\,,
\ee

\no and similarly for the tilde fields. Following the approach given in Ref. \cite{O}, the corrected expression for the Fermi thermofield is given by (see Appendix )

$$
\psi (x^\pm ; \beta )\,=\,\frac{1}{\sqrt{2 \pi}}\,\int_0^\infty dp \,\Big \{\,f_p (x^\pm)\,\Big ( b (\mp p)\,\cos \theta_F ( p ; \beta )\,+\,i\,\widetilde b^\dagger (\mp p)\,\sin \theta_F ( p ; \beta )\,\Big )\,
$$

\be
+\,
f^\ast_p (x^\pm)\,\Big ( d^\dagger (\mp p)\,\cos \theta_F ( p ; \beta )\,-\,i\,\widetilde d (\mp p)\,\sin \theta_F ( p ; \beta )\,\Big )\,\Big \}\,,
\ee

$$
\widetilde\psi (x^\pm ; \beta )\,=\,\frac{1}{\sqrt{2 \pi}}\,\int_0^\infty dp\, \Big \{\,f^\ast_p (x^\pm)\,\Big ( \widetilde b (\mp p)\,\cos \theta_F ( p ; \beta )\,-\,i\, b^\dagger (\mp p)\,\sin \theta_F ( p ; \beta )\,\Big )
$$

\be
+\,
f_p (x^\pm)\,\Big ( \widetilde d^\dagger (\mp p)\,\cos \theta_F ( p ; \beta )\,+\,i\, d (\mp p)\,\sin \theta_F ( p ; \beta )\,\Big )\,\Big \}\,,
\ee

\no where

\be
f_p (x)\,=\,e^{\,-\,i\,p\,x}\,.
\ee

\subsection{N-point function of chiral densities from fermionic point of view}

In order to compactify the calculation it proves convenient to introduce the following notation for the Fermi fields. We label them by an upper index $\eta$ taking the values zero and one, with the identifications 

$$
\psi^{\eta}=\psi\,,\mbox{ for $\eta=0$},
$$
\begin{equation}\label{tildefield-def}
\psi^{\eta}= {\tilde\psi}^\dagger\,, \mbox{for $\eta=1$}.
\end{equation}

\no Now, let us consider  the chiral densities of the massive free Fermi field in terms of the spinor components 

$$
\mathbf{J}^{\eta=0}_{+1}(x) = \dt {\psi}^\dagger_1(x)\psi_2(x) \dt\,,\quad \mathbf{J}^{\eta=0}_{-1}(x) = \dt {\psi}^\dagger_2(x)\psi_1(x) \dt\,,
$$
\begin{equation}\label{chiraldensities}
\mathbf{J}^{\eta=1}_{+1}(x) = \dt {\widetilde\psi}_1(x){\widetilde\psi}^\dagger_2(x) \dt\,,\quad \mathbf{J}^{\eta=1}_{-1}(x) \dt=  \dt {\widetilde\psi}_2(x){\widetilde\psi}^\dagger_1(x)\dt\,.
\end{equation}

\no In a compact notation these expressions become

\be
\mathbf J_+^\eta(x)= \dt {\psi^\eta_{1}}^\dagger\psi^\eta_{2} \dt\,\,\,,\,\,\,
\mathbf J_-^\eta(x)= \dt {\psi^\eta_{2}}^\dagger\psi^\eta_{1} \dt \,\,.
\ee

\no Here the notation for the lower index (chiral) has been chosen such as to facilitate later comparison with the corresponding bosonized expressions. In terms of the chiral densities the mass term reads,

\be
M (\overline{\psi}\psi - \overline{\widetilde\psi}\widetilde\psi ) =\,M\, \sum_{\lambda =\pm 1}\sum_{\eta =0,1}\mathbf{J}_\lambda^\eta\,.
\ee

\no Note that the minus sign on the left hand side for the tilded fields has been taken into account by reordering the fields on the rhs, taking account of the definitions (\ref{tildefield-def}) and of Fermi statistics. With the identification (\ref{LI}) for the interaction Lagrangian we have, setting $\mathbf{J} = \mathbf{J}^0$, 

$$
\langle 0(\beta)|T \mathbf{J}_{+1}(x_1)...\mathbf{J}_{+1}(x_{N})\mathbf{J}_{-1}(y_1)\cdots \mathbf{J}_{-1}(y_{N^\prime})|0(\beta)\rangle\,=
$$

\be\label{jstring}
\frac{1}{{\cal N}_F(\beta)}\quad \sum_{n=0}^{\infty}\frac{(-iM)^n}{n!}\sum_{(\lambda_k,\eta_k)}
\langle 0,\widetilde 0|T J_{+1}(x_1,\beta)...J_{+1}(x_N,\beta)J_{-1}(y_1,\beta)...J_{-1}(y_{N^\prime},\beta)
\int d{\cal Z}^n\prod_k^n \left( J_{\lambda_k}^{\eta_k}(z_k,\beta)\right)|\widetilde 0,0\rangle\,,
\ee

\no where we transfered the temperature dependence of the ground state to the densities by commuting the time independent unitary operator $U_F [\theta_F ]$ through the time ordering operation. $J_{\pm 1} (x ; \beta)$ are the thermal chiral densities of the free massless Fermi field

\be
J_{+ 1} (x ; \beta ) = U_F [\theta_F] J_{+ 1} (x) U_F^{- 1} [\theta_F ] = \dt \psi_1^\dagger (x^+ ; \beta) \psi_2 (x^- ; \beta ) \dt\,,
\ee

\be
J_{- 1}(x ; \beta ) = U_F [\theta_F] J_{- 1} (x) U_F^{- 1} [\theta_F ] = \dt \psi_2^\dagger (x^- ; \beta) \psi_1 (x^+ ; \beta ) \dt\,,
\ee

\no the integral $\int d{\cal Z}^n$ is short hand for

\be
\int d{\cal Z}^n = \prod_{k = 1}^n \int_{-\infty}^\infty d^2z_k\,,
\ee

\no and ${\cal N}_F (\beta )$  represents the contribution of the vacuum graphs,

\be
{\cal N}_F (\beta ) = \sum_{n=0}^{\infty}\frac{(-iM)^n}{n!}\sum_{(\lambda_k,\eta_k)}
\int d{\cal Z}^n \langle 0,\tilde 0|T\prod_k^n \left( J_{\lambda_k}^{\eta_k}(z_k,\beta)\right)|0,\tilde 0\rangle\,.
\ee

\no  We can thus apply Wick´s theorem to compute each term in the expansion in powers of $M$. The two-point functions for the fermion thermofield  have been computed in Ref.\cite{ABR} and adopting
the notation given in (\ref{tildefield-def}) the fermionic propagators are given by 

\be
\langle 0, \widetilde 0 \vert T \psi^0 (x^\pm ; \beta ) 
{\psi^0}^\dagger (y^\pm ; \beta ) \vert 0 , \widetilde 0
\rangle\, =
\frac{1}{2 i\,\beta\,\sinh [\frac{\pi}{\beta} (x^\pm - y^\pm-i\,\varepsilon\epsilon(x^0-y^0) )]}\,,
\ee

\be
\langle 0, \widetilde 0 \vert T {\psi^1} (x^\pm ; \beta ) 
{\psi^1}^\dagger (y^\pm ; \beta ) \vert 0 , \widetilde 0
\rangle\, =\,
\frac{-\,1}{2 i\,\beta\,\sinh [\frac{\pi}{\beta} (x^\pm - y^\pm \,+ \,i\,\varepsilon\epsilon(x^0-y^0) )]}\,,
\ee

\no and

\be
\langle 0, \widetilde 0 \vert T {\psi^0} (x^\pm ; \beta ) 
 {\psi^1}^\dagger (y^\pm ; \beta ) \vert 0 , \widetilde 0
\rangle\, =\,-\,\langle 0, \widetilde 0 \vert T {\psi^1} (x^\pm ; \beta ) 
 {\psi^0}^\dagger (y^\pm ; \beta ) \vert 0 , \widetilde 0
\rangle\,=\,
\frac{-\,i\,}{2 \,\beta\,\cosh [ \frac{\pi}{\beta} (x^\pm - y^\pm )]}\,.
\ee

\no These functions can be collected in a compact notation by using

\be
\cosh(x)=\pm i \sinh( x \mp i \pi/2)
\ee

\no so that we can write

\be\label{2pftilde}
\langle 0, \widetilde 0 \vert T \psi^{\eta_i} (x_i^\pm ; \beta ) 
{\psi^{\eta_j}}^\dagger (y_j^\pm ; \beta ) \vert 0 , \widetilde 0
\rangle\, =\,
\frac{( i)^{\eta_i + \eta_j}}{2 i\,\beta\,\sinh \left[\frac{\pi}{\beta} \left(x_i^\pm - i\frac{\beta\eta_i}{2} - y_j^\pm+i\frac{\beta\eta_j}{2}-i\,\varepsilon (- 1)^{\eta_j}\epsilon(x_i^0 - y_j^0) \right)\right]}\,. 
\ee

In applying the Wick's theorem only terms with equal number of $\psi_1^\eta$ and
$\psi_1^{\eta^\dagger}$ survive irrespective of their $\eta$ values. The same is true for the second spinor component. 
In terms of the chiral densities $J_\lambda^\eta$ this leads to the selection rule requiring  the sum of all values of $\lambda$ have to vanish, irrespective of the $\eta$ values.
There are thus equal number of positive and negative values of $\lambda$. We shall denote the space-time coordinates of the fields associated to $\lambda=1$ by $x_i$, while the ones associated to $\lambda=-1$ are denoted by $y_j$. The values of the $\eta$ upperscript are
accordingly splited as $\eta_i$ and $\eta_j$.  The result of the computation can be
written as

$$
\langle 0, \widetilde 0 \vert T \prod_{i=1}^{n} J_{\lambda_i = + 1}^{\eta_i} (x_i ; \beta )
\prod_{j = 1}^{n} J_{\lambda_j = - 1}^{\eta_j} (y_j ; \beta )
\vert 0,\widetilde 0\rangle =
 $$
 \begin{equation}\label{determinantalexpression}
 \det {\frac{( i )^{\eta_i + \eta_j}}{ \omega ((x^-_i-i \eta_i\beta/2)-(y^-_j-i \eta_j\beta/2))}}\det {\frac{( i )^{\eta_i + \eta_j}}{ \omega ((x^+_i-i \eta_i\beta/2)-(y^+_j-i \eta_j\beta/2))}},
 \end{equation}

\no where

\be\label{w}
\omega(x_i - y_j)= (2i\beta)\;\sinh \frac{\pi}{\beta}\big ( x_i - x_j \,-\,i (-1)^{\eta_j}\varepsilon \epsilon (x_i^0 - y_j^0) \big )\,. 
\ee

\no Note that the factors $( i )^{\eta_i + \eta_j}$ can be factorized out of the determinant computation
resulting in an overall sign depending on the total numbers of tilde fields $\psi_1$ and of corresponding tilde fields $\psi_2$. Both contributions lead to a factor $( - 1 )^{\widetilde m}$, with 

\be
\widetilde m = \sum_i^n \eta_i + \sum_j^n \eta_j
\ee

\no representing the total number of $\widetilde J_\pm$ in the string. 

Now using the factorization formula \cite{Mintchev} for the determinant of the $n \times n$ matrix
$1/\omega (x_i - y_j)$ 

\be
\det \frac{1}{\omega(x_i - y_j)}= \displaystyle 
\frac{\displaystyle \prod_{i< i^\prime}\omega(x_i - x_{i^\prime})\prod_{j < j^\prime}\omega(y_j - y_{j^\prime})}
{\displaystyle\prod_{i,j}\omega(x_i-y_j)}\,,
\ee

\no and suppressing for a while the $i \epsilon$ prescription, we obtain

$$
\langle 0, \widetilde 0 \vert T \prod_{i=1}^{n} J_{+1}^{\eta_i} (x_i)\prod_{j=1}^{n} J_{-1}^{\eta_j} (y_j)\vert
 0,\widetilde 0 \rangle =\quad
$$
$$
 (-1)^{\widetilde m}\,\,\frac{\displaystyle\prod_{i<i^\prime}^{n}\omega ((x^+_i - i \eta_i\beta/2)-(x^+_{i^\prime} - i \eta_{i^\prime} \beta/2))\,\displaystyle\prod_{j < j^\prime}^{n} \omega ((y^+_j - i \eta_j \beta/2)-(y^+_{j^\prime} - i \eta_{j^\prime}\beta/2))}{\displaystyle\prod_{i,j}^n\omega ((x^+_i-i \eta_i\beta/2)-(y^+_j-i \eta_j\beta/2))}\times
$$
$$
\frac{\displaystyle\prod_{i<i^\prime}^n\omega ((x^-_i-i \eta_i\beta/2)-(x^-_{i^\prime} - i \eta_{i^\prime}\beta/2)) \prod_{j < j^\prime}^n \omega ((y^-_j-i \eta_j\beta/2)-(y^-_{j^\prime}-i \eta_{j^\prime}\beta/2))}{\displaystyle\prod_{i,j}^n\omega ((x^-_i-i \eta_i\beta/2)-(y^-_j-i \eta_j\beta/2))} =
 $$
\begin{equation}\label{determinantalexpression2}
\quad  (-1)^{\widetilde m}\,\frac{\displaystyle\prod_{i<i^\prime}^n\Omega ((x_i-i \eta_i\beta/2)-(x_{i^\prime}-i \eta_{i^\prime}\beta/2)) \prod_{j < j^\prime}^n \Omega ((y_j-i \eta_j\beta/2)-(y_{j^\prime}-i \eta_{j^\prime}\beta/2))}{\displaystyle\prod_{i,j}^n\Omega ((x_i-i \eta_i\beta/2)-(y_j-i \eta_j\beta/2))}.
\end{equation}

\noindent Here we introduced the function 
 
\be
\Omega(x)=\omega(x^+)\omega(x^-)\,,
\ee

Returning the $i\epsilon$ prescription means that in the denominator of eq. (\ref{determinantalexpression2}) the change should be performed
 
\be
\sinh(\frac{\pi}{\beta} (x^+_i-y^+_j))\sinh(\frac{\pi}{\beta} (x^-_i-y^-_j))\longrightarrow \sinh(\frac{\pi}{\beta}( x^+_i-y^+_j-i\varepsilon_j\epsilon(x^0_i-y^0_j)))\sinh(\frac{\pi}{\beta}( x^-_i-y^-_j-i\varepsilon_j\epsilon(x^0_i-y^0_j)))\,,
\ee
\noindent where $\varepsilon_j=(-1)^{\eta_j}\varepsilon$.

Returning now to the expression (\ref{jstring}) we reorganize the terms in the expansion (\ref{jstring}) according to chirality, 
with $n$ being the total number of internal currents with $\lambda_k=1$ and $n^\prime$ the same for $\lambda_l=-1$. To the former we associate the variables $z_k$ and to the last ones $z^\prime_l$. The contributions can now be collected into the expression

$$
\langle 0(\beta)|T \Big ( \prod_{i = 1}^N \mathbf{J}_{+1}(x_i) \prod_{j = 1}^{N^\prime} \mathbf{J}_{-1}(y_j) \Big )|0(\beta)\rangle\,=
$$
$$
\frac{1}{{\cal N}(\beta)}\quad \sum_{n + n^\prime = 0}^{\infty }\!\!\!\!\!^\ast\,\,\frac{(-iM)^{n+n^\prime}}{n!n^\prime !}\sum_{(\eta_k,\eta_l)}\int d{\cal Z}^n\int d{\cal Z}^{n^\prime}
$$
\be\label{jstring2}
\langle 0,\widetilde 0|T J_{+1}(x_1,\beta)...J_{+1}(x_N,\beta)J_{-1}(y_1,\beta)...J_{-1}(y_{N^\prime},\beta)
\prod_k^n \left( J_{+1}^{\eta_k}(z_k,\beta)\right) \prod_l^{n^\prime} \left( J_{-1}^{\eta_l}(z_l^\prime,\beta)\right)|\widetilde 0,0\rangle\,,
\ee

\no where the notation $\displaystyle\sum \,^\ast$ means that the sum obeys the chiral conservation condition that now reads

\be\label{srf}
N-N^\prime+n-n^\prime=0\,,
\ee

\no and $d{\cal Z}^n$ refers to integration over  the  variables $\{z_1,\cdots,z_n\}$. Each term of this expansion is obtained directly from (\ref{determinantalexpression2}) leading to

$$
\langle 0(\beta) \vert T \Big ( \prod_{i = 1}^N \mathbf{J}_{+1}(x_i) \prod_{j = 1}^{N^\prime} \mathbf{J}_{-1}(y_j) \Big ) \vert 0(\beta)\rangle\,=
$$

\be
\frac{1}{{\cal N}_F (\beta)}\quad \sum_{n + n^\prime = 0}^{\infty }\!\!\!\!\!^\ast\,\,\frac{(-iM)^{n+n^\prime}}{n!n^\prime!}\sum_{(\eta_k,\eta_j)}\int d{\cal Z}^n\int d{\cal Z}^{n^\prime}\,(-1)^{\tilde m}{\cal I}(x,y;z,z'),
\ee

\noindent with

$$
{\cal I}(x,y;z,z')=  \frac{\displaystyle\prod_{i<i^\prime=1}^N \Omega(x_i-x_{i^\prime}) \prod_{j<j^\prime=1}^{N^\prime} \Omega({y}_{j}-{y}_{j^\prime})}{\displaystyle\prod_{i = 1}^N\prod_{j = 1}^{N^\prime} \Omega(x_i-y_{j})}\nonumber\times
$$

$$
\frac{ \displaystyle\prod_{i=1}^N\prod_{k=1}^n \Omega(x_i-(z_k-i\eta_k\beta/2))}
{\displaystyle\prod_{i=1}^{N}\prod_{l=1}^{n} \Omega({x}_{i}-(z^\prime_{l}-i\beta\eta_{l}/2))}
\frac{\displaystyle\prod_{j=1}^{N^\prime}\prod_{l=1}^{n^\prime} \Omega({y}_{j}-({z^\prime}_{l}-i{\eta}_{l}\beta/2))}{\displaystyle\prod_{k=1}^n
\prod_{j=1}^{N^\prime} \Omega(y_{j} - (z_k - i\eta_k\beta/2))}\times
$$

\be\label{fermionic}
\frac{\displaystyle\prod_{k<{k^\prime=1}}^n \Omega((z_k-i\eta_k\beta/2)-(z_{k^\prime}-i\eta_{k^\prime}\beta/2))
\prod_{l<l^\prime=1}^{n^\prime} \Omega(({z^\prime}_{l}-i{\eta}_{l}\beta/2)-({z^\prime}_{l^\prime}
-i{\eta}_{l^\prime}\beta/2))}{\displaystyle\prod_{k=1}^{N}\prod_{l=1}^{N^\prime} \Omega((z_k-i\eta_k\beta/2)-({z^\prime}_{l}-i{\eta}_{l}\beta/2))}
\ee

It is interesting to note, that the shift in the argument of the two-point function involving tilde field, (\ref{2pftilde}), (\ref{determinantalexpression}),
can be understood in the context of the real time formalism as the tilde field living on
the lower branch of the integration contour localized at $Im (t)= - i \beta/2$ in the complex time plane. This feature appears further in expression (\ref{fermionic}) since to all integrated variables $z_k$ are associated fields
with $\eta$ labels and these labels are summed for $\eta=0$ and $\eta=1$. In the real
time formalism this corresponds to contributions from the upper branch, the lower branch
and functions from mixed branches.  This has been observed to be necessary to satisfy the KMS conditions \cite{U}.

Note that all  expressions in the numerator link currents of the same chirality, whereas all terms in the denominator link currents of opposite chirality. We thus associate with each coordinate $\xi_i$ the chirality $\lambda_i$. We further reorganize the terms within the products by separating the integration variables into those which belong to ordinary currents and tilde currents (i.e. $z_k$ and $\tilde z_k$, respectively), and denote the respective chiralities by $\lambda_\ell$ and $- \tilde\lambda_k$. The assignment of $- \tilde\lambda_k$ to the tilde currents chirality is due to the definition Eq. (\ref{chiraldensities}). With this we can write expression (\ref{fermionic}) in the following form, suitable for later comparison:

$$ 
\langle 0 (\beta ) \vert\, T \Big ( \prod_{i = 1}^N \mathbf{J}_{+ 1} (x_i)\,\prod_{j = 1}^{N^\prime}\mathbf{J}_{- 1} ( y_j) \Big ) \,\vert 0 (\beta ) \rangle\,=\,\frac{1}{{\cal N}_F (\beta )}\,\sum_{n = 0}^\infty\,(-\,i\,M)^n \,\times
$$

$$
\sum_{m , \widetilde m}\frac{\delta_{m + \tilde m, n}}{m ! \,\widetilde m !}\,
( - 1 )^{\tilde m}\,
\int\,\prod_{\ell = 1}^m\,{\sum_{\{\lambda_l,\tilde\lambda_k\}}\!\!}^* d^2 z_\ell\,\int\,\prod_{k = 1}^{\widetilde m}\,d^2 \tilde z_k\,\times
$$

$$
\frac{\displaystyle \prod_{i^\prime > i}^N\,\big [ \Omega(x_i - x_{i^\prime} ; \beta  ) \big ]\, 
\prod_{j^\prime > j}^{N^\prime}\,\big [\Omega(y_{j} - y_{j^\prime} ; \beta )\big ]}{\displaystyle \prod_{i = 1}^N\,\prod_{j = 1}^{N^\prime}\,\big [\Omega(x_i - y_{j} ; \beta )\big ]}\,\times
$$

$$
\prod_{i = 1}^N \prod_{\ell = 1}^m\,\big [\Omega( x_i - z_\ell ; \beta )\big ]^{\lambda_\ell  }\,\prod_{i = 1}^{N} \prod_{k = 1}^{\widetilde m}\,\big [\Omega(x_i - \tilde z_k - \frac{i \beta}{2} )\big ]^{\,-\,\tilde \lambda_{k} } \,\times
$$

$$
\prod_{j = 1}^{N^\prime} \prod_{\ell = 1}^m\,\big [\Omega( y_j - z_\ell ; \beta )\big ]^{\,-\,\lambda_\ell  }\,\prod_{j = 1}^{N^\prime} \prod_{k = 1}^{\widetilde m}\,\big [\Omega(y_j - \tilde z_k  - \frac{i  \beta}{2} )\big ]^{\,\tilde \lambda_{k} } \,\times
$$

\be\label{num4f}
\prod_{\ell^\prime > \ell}^m\,\big [\Omega(z_\ell - z_{\ell^\prime} ; \beta )\big ]^{ \lambda_\ell \lambda_{\ell^\prime} }
\,\prod_{k^\prime > k}^{\widetilde m}\,\big [\widetilde\Omega(\tilde z_k - \tilde z_{k^\prime} ; \beta )\big ]^{\tilde \lambda_k \tilde \lambda_{k^\prime} }\,\prod_{\ell}^m\,\prod_k^{\widetilde m}\,\big [\Omega(z_\ell  - \tilde z_k - \frac{i \beta}{2} )\big ]^{\,-\,\lambda_\ell \tilde\lambda_k },
\ee

\noindent where
\be \label{omega1}
\Omega \big ( x_i\,-\,x_j ; \beta \big )\, =\,\Big ( 2 i \beta \Big )^2\,\sinh \big [ \frac{\pi}{\beta} \big (x_i^+\,-\,x_j^+\,-\,i \varepsilon \epsilon (x_i^0 - x_j^0 ) \big )\big ]
\sinh \big [ \frac{\pi}{\beta} \big (x_i^-\,-\,x_j^-\,-\,i \varepsilon \epsilon (x_i^0 - x_j^0 ) \big ) \big ]\,,
\ee

\be\label{omega2}
\widetilde \Omega \big ( x_i\,-\,x_j ; \beta \big )\, =\,\Big ( 2 i \beta \Big )^2\,\sinh \big [ \frac{\pi}{\beta} \big ( x_i^+\,-\,x_j^+\,+\,i \varepsilon \epsilon (x_i^0 - x_j^0 ) \big )\big ]
\sinh \big [ \frac{\pi}{\beta} \big (x_i^-\,-\,x_j^-\,+\,i \varepsilon \epsilon (x_i^0 - x_j^0 ) \big )\big ]\,,
\ee

\be\label{omega3}
\Omega \big ( x_i\,-\,x_j \,-\,\frac{i \beta}{2} \big )\, =\,\big ( 2 i \beta \big )^2\,\sinh \big [ \frac{\pi}{\beta} \big ( x_i^+\,-\,x_j^+\,-\,\frac{i \beta}{2}  \big )\big ]
\sinh \big [ \frac{\pi}{\beta} \big ( x_i^-\,-\,x_j^-\,-\,\frac{i \beta}{2}\big ) \big ]\,,
\ee
 
 \noindent and the sum over the chiralities $\lambda_\ell$ and $\tilde\lambda_k$ respects the total chirality condition

\begin{equation}\label{slr1}
N-N^\prime+\sum_\ell\lambda_\ell-\sum_k\tilde\lambda_k=0.
 \end{equation}

This concludes the computation of the N-point function of the chiral densities in the fermionic version. In the next Section we shall consider the computation of these N-point function within the bosonized version of the theory.



\section{Thermofield Bosonization point of view}
\setcounter{equation}{0}

The next step is to compute the N-point function of the chiral densities of the massive free fermion theory from the thermofield bosonization point of view. Since at $T = 0$ the bosonized theory describing both the massive free fermion theory  and the massive Thirring model is the sine-Gordon theory \cite{M,C} with  distinct values of the sine-Gordon parameter $\kappa$ ( different values of the scale dimension of the mass operator ), we shall consider the perturbative the perturbative computation of the N-point function of chiral densities of the bosonized  massive Thirring model from thermofield dynamics point of view. 

To begin with, let us consider the sine-Gordon theory for the doublet ($\Phi , \widetilde\Phi$), whose Lagrangian can be decomposed as

\be
\widehat{\cal L} (x) = \widehat{\cal L}^{(0)} (x)\,+\,\widehat{\cal L}^{I} (x)\,,
\ee

\no with

\be
\widehat{\cal L}^{(0)} (x) = {\cal L}^{(0)} (x) - \widetilde{\cal L}^{(0)} (x) = \frac{1}{2}\,\dt \partial_\mu \Phi (x)\,\partial^\mu \Phi (x)\,\dt\,-\,
\frac{1}{2}\,\dt \partial_\mu \widetilde\Phi (x)\,\partial^\mu \widetilde\Phi (x)\,\dt\,,
\ee

\be
\widehat{\cal L}_I (x)\,=\, M\,\Big ( \frac{\mu}{\pi} \Big )\,\Big ( \dt \cos \big (\kappa\,\Phi (x)\big ) \dt\,-\,\dt \cos \big (\kappa\,\widetilde\Phi (x)\big ) \dt \Big )\,,
\ee

\no where $\mu$ is the infrared regulator (IR) reminiscent of the free massless scalar theory. The scale dimension $D$ of the mass operator is given in terms of the Thirring coupling parameter $g$ as \cite{C}

\be
D\,=\,\frac{\kappa^2}{4 \pi}\,.
\ee

\no For the values $\kappa^2 = 4 \pi$, the sine-Gordon theory describes the massive free fermion theory discussed in the preceding section.

Now, let us introduce the Mandelstam representation \cite{M,RS} for the Fermi field operator

\be\label{M1}
\Psi (x)\, =\,\Big ( \frac{\mu}{2 \pi} \Big )^{\frac{1}{2}}\,
e^{\displaystyle - i\, \frac{\pi}{4}\,\gamma^5}\,
\,\dt e^{\textstyle\,i \frac{\kappa}{2} \gamma^5 \Phi (x) + 
\,i\,\frac{2 \pi}{\kappa}\,\displaystyle \int_{x^1}^\infty\,\partial_0 \Phi (x^0, z^1) 
dz^1\,}\dt\,,
\ee

\be\label{M2}
i\,\widetilde \Psi (x)\, =\,\Big ( \frac{\mu}{2 \pi} \Big )^{\frac{1}{2}}\,
e^{ \displaystyle - i\, \frac{\pi}{4}\,\gamma^5}\,
\,\dt e^{\textstyle\,i\,\frac{\kappa}{2} \gamma^5 \widetilde \Phi (x) + i \frac{2 \pi}{\kappa}  
\,\displaystyle \int_{x^1}^\infty\,\partial_0 \widetilde\Phi (x^0, z^1) 
dz^1 \big \}} \dt\,.
\ee

\no Notice that the field operator $\widetilde\Psi$ is not obtained from $\Psi$ by the ``tilde'' conjugation operation. This follows from the fact that $\widetilde\Psi$ should be an identical copy of $\Psi$ carrying the same charge and chirality quantum numbers. This is explained in detail in Section 4 and in the Appendix. 

The bosonized chiral densities $\mathbf{J}_{\pm 1} (x)$ are given by

\be\label{Jma}
\mathbf{J} _{+ 1} (x)\,=\,-\,\Big ( \frac{\mu}{2 \pi} \Big )\,\mathbf{W} (x)\,,
\ee

\be\label{Jme}
\mathbf{J}_{ - 1}(x)\,=\,-\,\Big (\frac{\mu}{2 \pi} \Big )\,\mathbf{W}^\ast (x)\,,
\ee

\no where the Wick-ordered exponentials carrying opposite chirality are

\be
\mathbf{W} (x)\,=\,\dt e^{i\,\kappa\,\Phi (x)}\dt\,.
\ee

\be
\mathbf{W}^\ast (x)\,=\,\dt e^{\,-\,i\,\kappa\,\Phi (x)}\dt\,.
\ee


\no The interaction picture vacuum is now given by

\be
\vert 0 , \beta \rangle\,=\,U_B [\theta_B ] \vert \widetilde 0 , 0 \rangle\,,
\ee

\no where the unitary operator taking one to the thermofields is given by ($\vert p^1 \vert = p$)

\be
U_B [\theta_B ]\,=\,e^{\,-\,\int_{- \infty}^{+ \infty} d p^1 \Big ( \tilde a (p^1) a (p^1)\,-\,a^\dagger (p^1) \tilde a^\dagger (p^1) \Big ) \theta_B (p , \beta ) }\,,
\ee

\no and the Bogoliubov parameter $\theta_B (p , \beta )$ is implicitly defined by

\be
\sinh \theta_B (p , \beta )\,=\,\displaystyle\frac{\displaystyle e^{\,-\,\frac{\beta p}{2}}}{\displaystyle\sqrt{1 - e^{\,-\,\beta p}}}\,,
\ee

\be
\cosh \theta_B (p , \beta )\,=\,\frac{1}{\sqrt{1 - e^{\,-\,\beta p}}}\,,
\ee

\no with the Bose-Einstein statistical weight given by

\be
N_B (p , \beta )\,=\,\sinh^2 \theta_B (p , \beta )\,=\,\frac{1}{e^{\,\beta p} - 1}\,.
\ee

\no Following the same procedure of the preceding section, in the bosonized theory the N-point function of the chiral densities is given by

$$
\langle 0 (\beta ) \vert\, T \Big ( \prod_{i = 1}^N \mathbf{J}_{+1} (x_i)\,\prod_{j = 1}^{N^\prime} \,\mathbf{J}_{-1} (y_{j}) \Big ) \,\vert 0 (\beta ) \rangle\,=
$$

\be\label{2pf}
\frac{\langle 0 , \widetilde 0 \vert T \Big ( \displaystyle\prod_{i = 1}^N J_{+1} (x_i ; \beta )\,\displaystyle\prod_{j = 1}^{N^\prime} \,
J_{-1} (y_{j} ; \beta ) 
\,\exp\,\Big \{ i\,\int\,\widehat{\cal L}_I ( z ; \beta )\,d \,^2 z \Big \}\Big )\vert \widetilde 0 , 0 \rangle}{\displaystyle\langle 0 , \widetilde 0 \vert T \exp\,\Big \{ i\,\int\, \widehat{\cal L}_I ( z ; \beta )\, d\,^2 z\Big \}\vert \widetilde 0 , 0 \rangle}\,,
\ee

\no where the interaction Lagrangian (in the interaction picture ) at finite temperature is

\be
\widehat{\cal L}_I (z ; \beta )\,=\,U_B [ \theta_B ] \widehat {\cal L}_I (z) U_B^{- 1} [ \theta_B ]\,=\,
\, M\,\Big (\frac{\mu}{ \pi}\,e^{\,-\,2\,D\,  \z (\beta , \mu^\prime)}\Big )\,\Big ( \dt \cos \big (\kappa\,\phi (z ; \beta )\big ) \dt\,-\,\dt \cos \big (\kappa\,\widetilde\phi (z ; \beta )\big ) \dt \Big )\,,
\ee

\no $J_{\pm 1} (x ; \beta )$ are the thermal chiral densities  in the interaction picture \cite{ABR}

\be
J_{+ 1} (x_i ; \beta )\,=\,U_B [ \theta_B ] J_{+ 1} (x_i) U_B^{- 1} [ \theta_B ]\,=\,-\,\Big (\frac{\mu}{2 \pi}\,e^{\,-\,2\,D \,\mathit{z}\, (\beta , \mu^\prime )} \Big )\,W ( x_i ; \beta , \lambda_{x_i} )\,,
\ee

\be
J_{- 1} ( y_j ; \beta )\,=\,\,U_B [ \theta_B ] J_{- 1} (y_j) U_B^{- 1} [ \theta_B ]\,=\,-\,\Big (\frac{\mu}{2 \pi}\,e^{\,-\,2\,D \,\mathit{z}\, (\beta , \mu^\prime )}\Big )\,{W^\ast} (  y_j ; \beta ,  \lambda_{y_j} )\,,
\ee

\no with the thermal Wick-ordered exponentials carrying opposite chirality given by ( in our convention $\lambda_{x_i} = \lambda_{y_j} = 1$)

\be
W (x_i ; \beta , \lambda_{x_i} )\,=\, \dt\,e^{\,i\,\kappa\,\lambda_{x_i}\,\phi (x_i ; \beta )}\,\dt\,,
\ee

\be
W^\ast (y_j ; \beta , \lambda_{y_j} )\,=\, \dt\,e^{\,-\,i\,\kappa\,\lambda_{y_j}\,\phi (y_j ; \beta )}\,\dt\,,
\ee

\no $\phi (x ; \beta )$ is a free massless pseudo-scalar thermofield \cite{ABR} and $\z (\beta , \mu^\prime )$ is the infrared divergent integral \cite{ABR}

\be
\z (\beta , \mu^\prime )\,=\,\int_{\mu^\prime}^\infty\,\frac{dp}{p ( e^{\beta p} - 1)}\,.
\ee

\no  Expanding the exponential in powers of $M$, introducing the tilde Wick-ordered exponential

\be
\widetilde W^\ast (z ; \beta , \lambda_z )\,=\,\dt e^{\,i\,\kappa\,\lambda_z\,\widetilde \phi (z ; \beta ) } \dt\,,
\ee

\no and using the fact that the fields $\phi (x ; \beta )$ and $\widetilde\phi (y ; \beta )$ commute \cite{ABR},  we get

$$
e^{\displaystyle i \int d^2 z \,\widehat{\cal L} (z) }\,=\,e^{\,\textstyle \,\,i\,M\,\Big (\frac{\mu}{ \pi}  \,e^{-\,2\,D \,\mathit{z}\,}\Big )\, \{\,\displaystyle \int d^2 z\,\dt\,\cos \kappa \phi (z ; \beta )\,\dt\,-\,\int d^2 \tilde z\,\dt\,\cos \kappa \widetilde\phi ( \tilde z ; \beta )\,\dt\,\Big \}}\,=\,
$$

$$
\sum_{n = 0}^\infty\,\frac{(\,i\,M)^n}{(n !)}\,\Big (\frac{\mu}{\pi} e^{-\,2\,D \,\mathit{z}}\Big )^{n}\,
\Big (\,\int d^2 z\,\dt\,\cos \kappa \phi (z ; \beta )\,\dt\,-\,\int d^2 \tilde z\,\dt\,\cos \kappa \widetilde\phi (\tilde z ; \beta )\,\dt\,\Big )^n=
$$

$$
\sum_{n = 0}^\infty\,\frac{(i\,M)^n}{(n !)}\,\Big (\frac{\mu}{ \pi}\,e^{-\,2\,D \,\mathit{z}}\Big )^{n } \,\sum_{m , \widetilde m}\frac{(n !)\,\delta_{m + \tilde m, n}}{m ! \,\widetilde m !}\,
( - 1 )^{\tilde m}\,\Big (\,\int d^2 z\,\dt\,\cos \kappa \phi (z ; \beta )\,\dt\,\Big )^m\,\Big (\,\int d^2 \tilde z\,\dt\,\cos \kappa \widetilde\phi (\tilde z ; \beta )\,\dt\,\Big )^{\widetilde m} =
$$

$$
\sum_{n = 0}^\infty\,(i\,M)^n \Big ( \frac{\mu}{ \pi } \,e^{\,-\,2\,D\,\mathit{z} }\Big )^{n}
\,\sum_{m , \widetilde m}\frac{\delta_{m + \tilde m, n}}{m ! \,\widetilde m !}\,
( - 1 )^{\tilde m}\,\int\,\prod_{\ell = 1}^m\, d^2 z_\ell\,\int\,\prod_{k = 1}^{\widetilde m}\,d^2 \tilde z_k\,\prod_{\ell = 1}^m\,\dt\,\cos \kappa \phi (z_\ell ; \beta )\,\dt\,\prod_{k = 1}^{\widetilde m}\,\dt\,\cos \kappa \widetilde\phi (\tilde z_k ; \beta )\,\dt
$$

$$
=\,\sum_{n = 0}^\infty\,(\,i\,M)^n \Big ( \frac{\mu}{2 \pi} e^{\,-\,2\,D \,\mathit{z} }\Big )^{n }
\,\sum_{m , \widetilde m}\frac{\,\delta_{m + \tilde m, n}}{m ! \,\widetilde m !}\,
( - 1 )^{\tilde m}\,\int\,\prod_{\ell = 1}^m\, d^2 z_\ell\,\int\,\prod_{k = 1}^{\widetilde m}\,d^2 \tilde z_k\,\times
$$

\be
\sum_{\{\lambda_\ell\}_{m}}\,\sum_{\{\tilde \lambda_k\}_{\tilde m}}\,
\prod_{\ell = 1}^m\,W (z_\ell ; \beta , \lambda_j )\,\prod_{k = 1}^{\widetilde m}\,\widetilde{W^\ast} (\tilde z_k ; \beta , \tilde\lambda_k )\,,
\ee

\no where $\lambda_\ell\,,\,\tilde\lambda_k\,=\,\pm\, 1$, and $\displaystyle\sum_{\{\lambda_\ell\}_{m}} (\displaystyle\sum_{\{\tilde\lambda_k\}_{\tilde m}})$ runs over all possibilities in the set $\{\lambda_1,\dots,\lambda_{m}\} (\{\tilde\lambda_1,\dots,\tilde\lambda_{\widetilde m}\})$. Denoting by ${\cal N}_B (\beta )$ the normalization factor, the Green's function (\ref{2pf}) can be written as

$$ 
\langle 0 (\beta ) \vert\, T \Big ( \prod_{i = 1}^N \mathbf{J}_{+ 1} (x_i)\,\prod_{j = 1}^{N^\prime}\mathbf{J}_{- 1} ( y_j) \Big ) \,\vert 0 (\beta ) \rangle\,=
$$

$$
\frac{1}{{\cal N}_B (\beta )}\,\sum_{n = 0}^\infty\,\big ( - 1 \big )^{(N + N^\prime)} (\,i\,M)^n\,\Big (\frac{\mu}{ 2 \pi}\,e^{\,-\,2\,D  \,\mathit{z} (\beta , \mu^\prime)}\Big )^{ (n + N + N^\prime)}\,
\sum_{m , \widetilde m}\frac{\delta_{m + \tilde m, n}}{m ! \,\widetilde m !}\,
( - 1 )^{\tilde m}\,\int\,\prod_{\ell = 1}^m\, d^2 z_\ell\,\int\,\prod_{k = 1}^{\widetilde m}\,d^2 \tilde z_k\,\times
$$

\be\label{num}
\sum_{\{\lambda_\ell \}_{m}}\,\sum_{\{\tilde \lambda_k\}_{\tilde m}}\,
\langle 0 , \widetilde 0 \vert \,T \Big (\prod_{i = 1}^N W (x_i ; \beta , \lambda_{x_i} )\,\prod_{j = 1}^{N^\prime} {W^\ast} (y_j) ; \beta , \lambda_{y_j} )\,\prod_{l = 1}^m\,W ( z_l ; \beta , \lambda_l )\,\prod_{k = 1}^{\widetilde m}\,\widetilde{W^\ast} ( \tilde z_k ; \beta , \tilde \lambda_k )\,\Big )\,\vert \widetilde 0 , 0 \rangle\,.
\ee

\no The time-ordered product of two Wick-ordered exponentials is defined by,

$$
T \Big ( W ( x ; \beta , \lambda_x )\, W (y ; \beta , \lambda_y )\,\Big )\,=
$$

$$
\,\dt \,W ( x ; \beta , \lambda_x )\, W (y ; \beta , \lambda_y )\,\dt\,\Big (\,\Big \langle W ( x ; \beta , \lambda_x ) W (y ; \beta , \lambda_y ) \Big \rangle \theta (x^0 - y^0)\,+\,
\Big \langle W ( y ; \beta , \lambda_y ) W (x ; \beta , \lambda_x ) \Big \rangle \theta (y^0 - x^0) \Big )
$$

\be
\equiv\,\dt \,W ( x ; \beta , \lambda_x )\, W (y ; \beta , \lambda_y )\,\dt\,\Big \langle T \big ( W ( x ; \beta , \lambda_x ) W (y ; \beta , \lambda_y ) \big ) \Big \rangle \,,
\ee

\no where

\be
\Big \langle W ( x ; \beta , \lambda_x ) W (y ; \beta , \lambda_y ) \Big \rangle\,=\,e^{\,-\,\lambda_x \lambda_y\,{\kappa}^2\,\langle 0 , \widetilde 0 \vert \phi (x ; \beta ) \phi (y ; \beta ) \vert \widetilde 0 , 0 \rangle }\,.
\ee

\no Using the identity ($\phi (x_i) = \phi_i$, $\theta (x_i^0 - y_j^0) = \theta_{ij}$)

\be
e^{\langle \phi_i \phi_j \rangle}\theta_{ij}\,+\,e^{\langle \phi_j \phi_i \rangle}\theta_{ji}\,\equiv\,
e^{\langle \phi_i \phi_j \rangle \theta_{ij}\,+\,\langle \phi_j \phi_i \rangle \theta_{ji}}\,=\,e^{\langle T \phi_i \phi_j \rangle }\,,
\ee

\no we can write

\be
T \Big ( W ( x ; \beta , \lambda_x )\, W (y ; \beta , \lambda_y )\,\Big )\,=
\,\dt \,W ( x ; \beta , \lambda_x )\, W (y ; \beta , \lambda_y )\,\dt\,e^{-\,\lambda_x \lambda_y \,{\kappa}^2\,\langle T \phi (x ; \beta ) \phi (y ; \beta ) \rangle }\,. 
\ee

\no The Wick's theorem can be extended to generalized time-ordered product of Wick-ordered exponentials and we obtain  

$$
\Big \langle T \Big ( \prod_{j = 1}^\ell W (x_j ; \beta , \lambda_j ) \prod_{k = 1}^{\tilde\ell} \widetilde W (\tilde x_k ; \beta , \tilde\lambda_k ) \Big ) \Big \rangle \,=\,
$$

\be
\prod_{j > j^\prime}^\ell \Big \langle T \big ( W (x_j ; \beta , \lambda_j ) W (x_{j^\prime} ; \beta , \lambda_{j^\prime} ) \big ) \Big \rangle\,
\prod_{k > k^\prime}^{\tilde\ell} \Big \langle T \big (\widetilde W (\tilde x_k ; \beta , \tilde\lambda_{k} ) \widetilde W (\tilde x_{k^\prime} ; \beta , \tilde\lambda_{k^\prime} ) \big ) \Big \rangle\,
\prod_{j , k }^{\ell , \tilde\ell}  \Big \langle T \big ( W (x_j ; \beta , \lambda_j ) \widetilde  W (\tilde x_k ; \beta , \tilde\lambda_k ) \big ) \Big \rangle\,.
\ee

\no In this way  one gets for the Fock vacuum expectation value of $T$-ordered product of Wick exponentials in (\ref{num}),

$$
\langle 0 , \widetilde 0 \vert \,T \Big ( \prod_{i = 1}^N W (x_i ; \beta , \lambda_i )\,\prod_{j = 1}^{N^\prime} {W^\ast} ( y_j ; \beta , \lambda_j )\,\prod_{\ell = 1}^m\,W ( z_\ell ; \beta , \lambda_\ell )\,\prod_{k = 1}^{\widetilde m}\,\widetilde{W^\ast} ( \tilde z_k ; \beta , \tilde \lambda_k )\,\Big )\,\vert \widetilde 0 , 0 \rangle\,=
$$

$$
\prod_{i^\prime > i}^N\,\Big \langle T \big ( W (x_i ; \beta , \lambda_i )W (x_{i^\prime} ; \beta , \lambda_{i^\prime} ) \big ) \Big \rangle 
\prod_{j^\prime > j}^{N^\prime}\,\Big \langle T \big ({W^\ast} (y_{j} ; \beta ,  \lambda_j ) {W^\ast} 
(y_{j^\prime} ; \beta ,  \lambda_{j^\prime} ) \big ) \Big \rangle \times
$$

$$\prod_{i = 1}^N\,\prod_{j = 1}^{N^\prime}\,\Big \langle T \big ( W (x_i ; \beta , \lambda_i ) {W^\ast} (
y_{j} ; \beta , \lambda_j ) \big ) \Big \rangle \times 
$$

$$
\prod_{i = 1}^N \prod_{\ell = 1}^m\,\Big \langle T \big ( W (x_i ; \beta , \lambda_i ) W (z_\ell ; \beta , \lambda_\ell ) \big ) \Big \rangle\,\prod_{i = 1}^{N} \prod_{k = 1}^{\widetilde m}
\Big \langle T \big ( {W} (x_i ; \beta , \lambda_i ) \widetilde W^\ast (\tilde z_k ; \beta , \tilde \lambda_k ) \big ) \Big \rangle\,\times
$$

$$
\prod_{j= 1}^{N^\prime} \prod_{\ell = 1}^{m}\,\Big \langle T \big ( W^\ast (y_j ; \beta , \lambda_j ) {W} ( z_\ell ; \beta ,  \lambda_\ell ) \big ) \Big \rangle\,
\prod_{j= 1}^{N^\prime} \prod_{k = 1}^{\widetilde m} \Big \langle T \big ( {W^\ast} ( y_j ; \beta , \lambda_j ) \widetilde{W^\ast} (\tilde z_k ; \beta , \tilde \lambda_k ) \big ) \Big \rangle\,\times
$$

$$
\prod_{\ell^\prime > \ell}^m\,\Big \langle T \big ( W (z_\ell ; \beta , \lambda_\ell ) W (z_{\ell^\prime} ; \beta , \lambda_{\ell^\prime} ) \big ) \Big \rangle\,\prod_{k^\prime > k}^{\widetilde m}\,
\Big \langle T \big ( \widetilde{W^\ast} (\tilde z_k ; \beta , \tilde \lambda_k ) \widetilde{W^\ast} (\tilde z_{k^\prime} ; \beta ,  \tilde\lambda_{k^\prime} ) \big ) \Big \rangle\,\times
$$

\be\label{num2}
\prod_{\ell}^m\,\prod_k^{\widetilde m}\,\Big \langle T \big ( W (z_\ell ; \beta , \lambda_\ell ) \widetilde{W^\ast} (\tilde z_k ; \beta , \tilde \lambda_k ) \big ) \Big \rangle\,.
\ee

\no The propagators of the scalar thermofields can be written in terms of the propagators of the Fermi thermofields as follows \cite{ABR}

\be\label{2pf1}
\langle 0 , \widetilde 0 \vert T \phi (x ; \beta ) \phi (y ; \beta ) \vert \widetilde 0 , 0 \rangle\,=\,\frac{1}{\pi}\,\mathit{z} (\mu^\prime , \beta )\,-\,\frac{1}{2 \pi} \ln \big (\frac{\mu}{2 \pi}\big )\,-\,\frac{1}{4 \pi}\,\ln \Omega \big (x - y ; \beta \big )\,,
\ee

\be\label{2pf2}
\langle 0 , \widetilde 0 \vert T \widetilde \phi (x ; \beta ) \widetilde \phi (y ; \beta ) \vert \widetilde 0 , 0 \rangle\,=\,
-\,\frac{i}{2}\,+\,\frac{1}{\pi}\,\mathit{z} (\mu^\prime , \beta )\,-\,\frac{1}{2 \pi} \ln \big (\frac{\mu}{2 \pi}\big )\,-\,\frac{1}{4 \pi}\,\ln \widetilde \Omega \big (x - y ; \beta \big )
\ee

\be\label{2pf3}
\langle 0 , \widetilde 0 \vert T \phi (x ; \beta ) \widetilde \phi (y ; \beta ) \vert \widetilde 0 , 0 \rangle\,=\,-\,\frac{1}{2\pi}\,f (\mu^{\prime\prime} , \beta )\,-\,\frac{1}{2 \pi}\,\ln \big (2 \beta \big )\,+\,\frac{1}{4 \pi}\,\ln \Omega \big (x - y -  \frac{i \beta}{2} \big )\,,
\ee

\no where the $\Omega$'s are defined in (\ref{omega1})-(\ref{omega3}). The dependence on the infrared cut-offs $\mu^\prime$ and $\mu^{\prime\prime}$ are given by \cite{ABR}

\be
\z (\mu^\prime , \beta )\,=\,\int_{\mu^\prime}^\infty\,\frac{dp}{p} (e^{\,\beta p}\,-\,1)^{- 1}\,,
\ee

\be
f (\mu^{\prime\prime} , \beta )\,=\,\int_{\mu^{\prime\prime}}^\infty\,\frac{dp}{p} \frac{1}{\sinh (\frac{\beta p}{2})}\,,
\ee

\no and the corresponding asymptotic behavior are 

\be\label{abz}
\z (\mu^\prime \approx 0 , \beta ) \rightarrow\,\frac{1}{\beta \mu^\prime}\,+\,\frac{1}{2}\,\ln (\beta \mu^\prime)\,,
\ee

\be\label{abf}
f (\mu^{\prime\prime} \approx 0 ,  \beta ) \rightarrow \frac{2}{\beta \mu^{\prime\prime}}\,.
\ee

\no The N-point function is then given by ($\lambda_{{x_i}} = \lambda_{{y_j}} = 1$)

$$ 
\langle 0 (\beta ) \vert\, T \Big ( \prod_{i = 1}^N \mathbf{J}_{+ 1} (x_i)\,\prod_{j = 1}^{N^\prime}\mathbf{J}_{- 1} ( y_j) \Big ) \,\vert 0 (\beta ) \rangle\,=\,\frac{1}{{\cal N}_B (\beta )}\,\sum_{n = 0}^\infty\,\big ( - 1 \big )^{(N + N^\prime)}\,(\,i\,M)^n \,\times
$$

$$
\sum_{m , \widetilde m}\frac{\delta_{m + \tilde m, n}}{m ! \,\widetilde m !}\,( - 1 )^{\tilde m}\,
\sum_{\{\lambda_\ell \}_{m}}\,\sum_{\{\tilde \lambda_k\}_{\tilde m}}\,F^{(m , \widetilde m)} (\mu , \mu^\prime , \mu^{\prime\prime} , \beta )\,G^{(m , \widetilde m)} (\mu , \mu^\prime , \beta )\,
\int\,\prod_{\ell = 1}^m\, d^2 z_\ell\,\int\,\prod_{k = 1}^{\widetilde m}\,d^2 \tilde z_k\,\times
$$

$$
\displaystyle\frac{\displaystyle \prod_{i^\prime > i}^N\,\big [ \Omega (x_i - x_{i^\prime} ; \beta  ) \big ]^{ D \lambda_i\lambda_{i^\prime} }\, 
\prod_{j^\prime > j}^{N^\prime}\,\big [\Omega (y_{j} - y_{j^\prime} ; \beta )\big ]^{D \lambda_j \lambda_{j^\prime}}}{\displaystyle\prod_{i = 1}^N\,\prod_{j = 1}^{N^\prime}\,\big [\Omega (x_i - y_{j} ; \beta )\big ]^{  D \lambda_i \lambda_j }}\,\times 
$$

$$
\prod_{i = 1}^N \prod_{\ell = 1}^m\,\big [\Omega ( x_i - z_\ell ; \beta )\big ]^{D \lambda_i \lambda_\ell  }\,\prod_{i = 1}^{N} \prod_{k = 1}^{\widetilde m}\,\big [\Omega (x_i - \tilde z_k - \frac{i \beta}{2} )\big ]^{\,-\,D \lambda_i \tilde \lambda_{ k} } \,\times
$$

$$
\prod_{j = 1}^{N^\prime} \prod_{\ell = 1}^m\,\big [\Omega ( y_j - z_\ell ; \beta )\big ]^{\, - \,D \lambda_j \lambda_\ell  }\,\prod_{j = 1}^{N^\prime} \prod_{k = 1}^{\widetilde m}\,\big [\Omega (y_j - \tilde z_k  - \frac{i  \beta}{2} )\big ]^{\, D \lambda_j \tilde \lambda_{k} } \,\times
$$

\be\label{num4}
\frac{\displaystyle\prod_{\ell^\prime > \ell}^m\,\big [\Omega (z_\ell - z_{\ell^\prime} ; \beta )\big ]^{ D \lambda_\ell \lambda_{\ell^\prime} }
\,\prod_{k^\prime > k}^{\widetilde m}\,\big [\widetilde \Omega (\tilde z_k - \tilde z_{k^\prime} ; \beta )\big ]^{D \tilde \lambda_k \tilde \lambda_{k^\prime} }}{\displaystyle \prod_{\ell}^m\,\prod_k^{\widetilde m}\,\big [\Omega (z_\ell  - \tilde z_k - \frac{i \beta}{2} )\big ]^{\,D \lambda_\ell \tilde\lambda_k }}\,K\,,
\ee

\no where the phase

\be
K\,=\,e^{\,\displaystyle 2 i \pi D \sum_{k^\prime > k}^{\widetilde m} \tilde\lambda_k \tilde\lambda_{k^\prime}}\,,
\ee

\no is the identity for integer values of the scale dimension $D$ and the cut-off dependence is  given by

\be\label{G}
G^{(m , \widetilde m)} (\mu , \mu^\prime,  \beta )\,=\,
\Big ( \frac{\mu}{\pi} \Big )^{\,(1\,-\,D) (N + N^\prime + n )}\,\Big[  \frac{\mu}{2 \pi} \,e^{\,-\,2\,\z (\mu^\prime ; \beta )}\,\Big ]^{\displaystyle  D \Big ( N \,-\,N^\prime\, + \,\sum_{\ell = 1}^m \lambda_\ell - \sum_{k = 1}^{\widetilde m} \tilde \lambda_k \Big )^2}\,,
\ee

$$
F^{(m , \widetilde m)} (\mu, \mu^\prime, \mu^{\prime\prime}, \beta )\,=\,\Big [ \Big (\frac{\beta \mu}{\pi}\Big )^{2 D} \,e^{ 2 D \big (f (\mu^{\prime\prime}, \beta )\,-\,2\,\z (\mu^\prime, \beta) \big )} \Big ]^{
 \displaystyle  \Big ( \sum_{k = 1}^{\widetilde m} \tilde \lambda_k \Big ) \Big (  N \, -\,{N^\prime}  + \sum_{\ell = 1}^m \lambda_\ell \Big )}
$$

\be\label{F}
=\,\Big [ \big ( \frac{\mu}{\pi \mu^\prime} \big )^{2 D }\,e^{\,\displaystyle \frac{4 D}{\beta} \Big ( \frac{1}{\mu^{\prime\prime}}\,-\,\frac{1}{\mu^\prime} \Big )} \Big ]^{ \displaystyle \Big ( \sum_{k = 1}^{\widetilde m} \tilde \lambda_k \Big ) \Big (  N \, -\,{N^\prime}  + \sum_{\ell = 1}^m \lambda_\ell \Big )}\,.
\ee

In arriving at these expressions we collected the cut-off dependent terms, noting that

$$
\sum_{i^\prime > i}^N \lambda_i \lambda_{i^\prime}
+ \sum_{j^\prime > j}^{N^\prime} \lambda_j \lambda_{j^\prime}
- \sum_{i}^N \lambda_i \sum_j^{N^\prime}\lambda_{j}
+ \sum_{i}^N \lambda_i \sum_\ell^m \lambda_\ell
- \sum_{j}^{N^\prime} \lambda_j \sum_\ell^m \lambda_\ell
+ \sum_{\ell^\prime > \ell}^m \lambda_\ell \lambda_{\ell^\prime}
+ \sum_{k^\prime > k}^{\widetilde m} \tilde\lambda_k \tilde\lambda_{k^\prime} =
$$

\be
\frac{1}{2}\,\Big ( N \,-\,N^\prime\, + \,\sum_{\ell = 1}^m \lambda_\ell - \sum_{k = 1}^{\widetilde m} \tilde \lambda_k \Big )^2\,-\,\frac{1}{2} \big ( N + N^\prime + n \big )\,+\,
\sum_k^{\widetilde m} \tilde \lambda_k \Big ( N - N^\prime\,+\,\sum_\ell^m \lambda_\ell \Big )\,,
\ee

\no with ($\lambda_i = \lambda_j = 1$)

\be
\sum_{i = 1}^N \lambda_{i} = \,N\,=\,\sum_{i = 1}^N (\lambda_i)^2 \,,
\ee

\be
\sum_{j = 1}^{N^\prime} \lambda_j = \,N^\prime\,=\,\sum_{j = 1}^{N^\prime} (\lambda_j)^2\,,
\ee

\no and made use in (\ref{F}) of the asymptotic behavior (\ref{abz}) and (\ref{abf}). 

Now, let us consider the free massive 
fermion theory ($D = 1$), which is an infrared cut-off independent scale non-invariant theory.

Following the procedure introduced in Ref. \cite{ABR} (see Appendix), the theory of the free massless scalar thermofields $\phi (x ; \beta  )$ and $\widetilde \phi (x ; \beta )$ can be constructed as the zero mass limit of the massive free scalar thermofields $\Sigma (x ; \beta )$ and $\widetilde\Sigma (x ; \beta )$. In this way, the infrared regulator $\mu$ of the zero temperature two-point function should be identified with the infrared cut-offs $\mu^\prime$ and $\mu^{\prime\prime}$ of the temperature-dependent contributions $\z (\mu^\prime , \beta )$ and $f (\mu^{\prime\prime} , \beta )$,

\be
\mu^{\prime\prime}\,=\,\mu^\prime\,=\,\frac{\mu}{\pi}\,,
\ee

\no such that

\be
F^{(m , \widetilde m)}(\mu , \mu^\prime , \mu^{\prime\prime} , \beta ) {\Big \vert}_{_{\,\mu^{\prime\prime} = \mu^\prime = \frac{\mu}{\pi}}}\,
\,=\,1\,.
\ee

\no In this way, the only non zero contributions in the expansion (\ref{num4}) are those which satisfy the selection rule

\be\label{SR}
 N \,-\,{N^\prime}  + \sum_{\ell = 1}^m \lambda_\ell - \sum_{k = 1}^{\widetilde m} \tilde \lambda_k \,=\,0\,.
\ee

The selection rule (\ref{SR}) is in agreement with our previous computations (Eq. (\ref{srf})) for the massive free fermionic theory ($D = 1$). Indeed, for $D = 1$ the bosonized version of the perturbative expansion given by (\ref{num4}) coincide with our previous result in the fermionic formulation obtained in Section 2. In order to recover the factor $( - 1 )^n$ that appears in (\ref{num4f}), we shall make use of the  selection rule (\ref{SR}). Taking into account the summations $\displaystyle \sum_{\{\lambda_\ell \}_{m}}\,\sum_{\{\tilde \lambda_k\}_{\tilde m}}$ in (\ref{num4}), one can write

$$
\big ( - 1 \big )^{\big ( N + N^\prime \big )} \big ( i M \big )^n\,=\,
\big ( - 1 \big )^{\big ( N + N^\prime + n \big )} \big ( - i M \big )^n\,\equiv\,
\big ( - 1 \big )^{\big ( N - N^\prime + n \big )} \big ( - i M \big )^n\,\equiv
$$

\be
\big ( - 1 \big )^{\displaystyle \big ( N - N^\prime + \sum_{\ell = 1}^m \lambda_\ell + \sum_{k = 1}^{\widetilde m} \tilde\lambda_k \big )} \big ( - i M \big )^n\,\equiv\,
\big ( - 1 \big )^{\displaystyle \big ( N - N^\prime + \sum_{\ell = 1}^m \lambda_\ell - \sum_{k = 1}^{\widetilde m} \tilde\lambda_k \big )} \big ( - i M \big )^n\,=\, \big ( - i M \big )^n\,.
\ee

\no In this way, for $D = 1$, the bosonized $N$-point function (\ref{num4}) can be written as

$$ 
\langle 0 (\beta ) \vert\, T \Big ( \prod_{i = 1}^N \mathbf{J}_{+ 1} (x_i)\,\prod_{j = 1}^{N^\prime}\mathbf{J}_{- 1} ( y_j) \Big ) \,\vert 0 (\beta ) \rangle\,=
$$

$$
\frac{1}{{\cal N}_B (\beta )}\,\sum_{n = 0}^\infty\,(\,- i\,M)^n \,\sum_{m , \widetilde m}\frac{\delta_{m + \tilde m, n}}{m ! \,\widetilde m !}\,( - 1 )^{\tilde m}\,
\sum_{\{\lambda_\ell \}_{m}}\,\sum_{\{\tilde \lambda_k\}_{\tilde m}}\,\int\,\prod_{\ell = 1}^m\, d^2 z_\ell\,\int\,\prod_{k = 1}^{\widetilde m}\,d^2 \tilde z_k\,\times
$$

$$
\displaystyle\frac{\displaystyle \prod_{i^\prime > i}^N\,\big [ \Omega (x_i - x_{i^\prime} ; \beta  ) \big ]\, 
\prod_{j^\prime > j}^{N^\prime}\,\big [\Omega (y_{j} - y_{j^\prime} ; \beta )\big ]}{\displaystyle\prod_{i = 1}^N\,\prod_{j = 1}^{N^\prime}\,\big [\Omega (x_i - y_{j} ; \beta )\big ]}\,\times 
$$

$$
\prod_{i = 1}^N \prod_{\ell = 1}^m\,\big [\Omega ( x_i - z_\ell ; \beta )\big ]^{\lambda_\ell  }\,\prod_{i = 1}^{N} \prod_{k = 1}^{\widetilde m}\,\big [\Omega (x_i - \tilde z_k - \frac{i \beta}{2} )\big ]^{\,-\,\tilde \lambda_{ k} } \,\times
$$

$$
\prod_{j = 1}^{N^\prime} \prod_{\ell = 1}^m\,\big [\Omega ( y_j - z_\ell ; \beta )\big ]^{\, - \, \lambda_\ell  }\,\prod_{j = 1}^{N^\prime} \prod_{k = 1}^{\widetilde m}\,\big [\Omega (y_j - \tilde z_k  - \frac{i  \beta}{2} )\big ]^{\, \tilde \lambda_{k} } \,\times
$$

\be\label{num4fb}
\prod_{\ell^\prime > \ell}^m\,\big [\Omega (z_\ell - z_{\ell^\prime} ; \beta )\big ]^{  \lambda_\ell \lambda_{\ell^\prime} }
\,\prod_{k^\prime > k}^{\widetilde m}\,\big [\widetilde \Omega (\tilde z_k - \tilde z_{k^\prime} ; \beta )\big ]^{ \tilde \lambda_k \tilde \lambda_{k^\prime} } \prod_{\ell}^m\,\prod_k^{\widetilde m}\,\big [\Omega (z_\ell  - \tilde z_k - \frac{i \beta}{2} )\big ]^{\,- \lambda_\ell \tilde\lambda_k }\,,
\ee

\no in agreement with  the corresponding function (\ref{num4f}) obtained in the original fermionic version of the theory. This proves the formal equivalence of the massive free fermion thermofield theory with the sine-Gordon thermofield model for the particular value of the sine-Gordon parameter $\kappa ^2 = 4 \pi$.

\section{Physical Meaning of the Selection Rule}
\setcounter{equation}{0}

In this section we shall discuss the physical interpretation of the bosonized expression (\ref{M2}) for the field $\widetilde\Psi$ and its connection with the interpretation of the selection rule (\ref{SR}). To begin with, let us consider the massless free fermionic theory described by the doublet $\pmatrix{ \psi \cr - i{\widetilde\psi}^\dagger }$. The corresponding bosonized expressions are given by (\ref{M1})-(\ref{M2}) with $\kappa = 2 \sqrt\pi$, and can be written as 

\be\label{pb}
\psi (x ) \,=\,\Big ( \frac{\mu}{2 \pi} \Big )^{\frac{1}{2}}\,\dt\,e^{\textstyle\, i \sqrt \pi\,\{ \gamma^5 \phi (x)\,+\,\varphi (x) \}}\,\dt\,,
\ee

\be\label{ptb}
i \widetilde \psi (x) \,=\,\Big ( \frac{\mu}{2 \pi} \Big )^{\frac{1}{2}}\,\dt\,e^{\textstyle\, i \sqrt \pi\,\{ \gamma^5 \widetilde\phi (x)\,+\,\widetilde\varphi (x) \}}\,\dt\,,
\ee

\no with $\epsilon_{\mu \nu} \partial^\nu \varphi\,=\,\partial_\mu \phi$. The bosonic doublet $(\phi , \widetilde\phi )$ is described by the total Lagrangian

\be
\widehat{\cal L}\,=\,\frac{1}{2} \big ( \partial_\mu \phi \big )^2\,-\,\frac{1}{2} \big ( \partial_\mu \widetilde\phi \big )^2\,.
\ee

\no The physical interpretation for the fact that the bosonized expression (\ref{ptb}) for the field $\widetilde\psi$ is not obtained from the corresponding bosonized expression (\ref{pb}) for the field $\psi$ by the ``tilde conjugation operation'' is the following: In the Thermofield Dynamics formalism, the fictitious ``tilde'' system should be an identical copy of the system under consideration, which implies that the field $\widetilde\psi$ should be an identical copy of $\psi$, i. e., carrying the same charge and chirality quantum numbers. At $T = 0$, we obtain the following equal-time commutation relations for the free scalar fields \footnote{The corresponding canonical momenta are

$$
\Pi (x)\,=\,\partial_0 \phi (x)\,,
$$

$$
\widetilde\Pi (x)\,=\,-\,\partial_0 {\widetilde\phi} (x)\,,
$$

\no in such a way that the canonical equal-time commutation relations are given by

$$
\big [ \phi (x)\,,\,\Pi (y) \big ]\,=\,i\,\delta (x^1\,-\,y^1)\,,
$$

$$
\big [ \widetilde \phi (x)\,,\,\widetilde\Pi (y) \big ]\,=\,i\,\delta (x^1\,-\,y^1)\,.
$$

\no The dynamical equations are

$$
\partial_0 \phi\,=\,-\,i\,\big [ \phi (x)\,,\,H \big ]\,=\,-\,i\,\big [ \phi (x)\,,\,\widehat H \big ]\,,
$$

$$
\partial_0 \widetilde\phi\,=\,i\,\big [ \widetilde\phi (x)\,,\,\widetilde H \big ]\,=\,-\,i\,\big [ \widetilde\phi (x)\,,\,\widehat H \big ]\,,
$$

where the total Hamiltonian $\widehat H\,=\,H\,-\,\widetilde H$ is the generator of time evolution of the combined system.}

\be\label{cr1}
\big [ \phi (x)\,,\,\partial_0 \phi (y) \big ]\,=\,i\,\delta (x^1\,-\,y^1)\,,
\ee

\be\label{cr2}
\big [ \widetilde \phi (x)\,,\,\partial_0 {\widetilde\phi} (y) \big ]\,=\,-\,i\,\delta (x^1\,-\,y^1)\,.
\ee

\no Within the Thermofield Dynamics approach the above algebraic relations are retained at finite temperature. The computation of the bosonized expression for the fermionic currents ${\cal J}^\mu = \overline\psi \gamma^\mu \psi$ and $\widetilde{\cal J}^\mu = \overline{\widetilde\psi} \gamma^\mu \widetilde\psi$ of the free massless theory were performed in Ref. \cite{ABR}. Using the bosonized expressions for the free massless fermion fields (\ref{pb})-(\ref{ptb}),  one finds

\be
{\cal J}_\mu (x ; \beta )\,=\,-\,\frac{1}{\sqrt\pi}\,\partial_\mu \varphi (x ; \beta)\,,
\ee

\be
\widetilde{\cal J}_\mu (x ; \beta )\,=\,+\,\frac{1}{\sqrt\pi}\,\partial_\mu \widetilde\varphi (x ; \beta)\,.
\ee

\no The axial-vector currents are ($\gamma^5 \gamma^\mu = \epsilon^{\mu \nu} \gamma_\nu$)

\be
{\cal J}^5_\mu (x ; \beta )\,=\,-\,\frac{1}{\sqrt\pi}\,\partial_\mu \phi (x ; \beta)\,,
\ee

\be
{\widetilde{\cal J}}^5_\mu (x ; \beta )\,=\,+\,\frac{1}{\sqrt\pi}\,\partial_\mu \widetilde\phi (x ; \beta)\,.
\ee

\no Introducing the corresponding charges

\be
{\cal Q} (\beta )\,=\,\int_{- \infty}^{+\infty} J_0 (z ; \beta ) d z ^1\,\,\,\,\,,\,\,\,\,\,
\widetilde{\cal Q} (\beta )\,=\,\int_{- \infty}^{+\infty} \widetilde J_0 (z ; \beta ) d z ^1\,,
\ee

\be
{\cal Q}^5 (\beta )\,=\,\int_{- \infty}^{+\infty} J_0^5 (z ; \beta ) d z ^1\,\,\,\,\,,\,\,\,\,\,
{\widetilde{\cal Q}} ^5 (\beta )\,=\,\int_{- \infty}^{+\infty} {\widetilde J_0} ^5 (z ; \beta ) d z ^1\,,
\ee

\no and using (\ref{cr1})-(\ref{cr2}) we get

\be
\big [ {\cal Q} (\beta )\,,\,\psi (x ; \beta ) \big ]\,=\,-\,\psi (x ; \beta )\,,
\ee

\be
\big [ {\cal Q}^5 (\beta )\,,\,\psi (x ; \beta ) \big ]\,=\,-\,\gamma^5\,\psi (x ; \beta )\,,
\ee

\be
\big [ \widetilde{\cal Q} (\beta )\,,\,\widetilde\psi (x ; \beta ) \big ]\,=\,-\,\widetilde\psi (x ; \beta )\,,
\ee

\be
\big [ \widetilde{\cal Q}^5 (\beta )\,,\,\widetilde\psi (x ; \beta ) \big ]\,=\,-\,\gamma^5\,\widetilde\psi (x ; \beta )\,,
\ee

\no implying that $\psi$ and $\widetilde\psi$ carry the same charge and chirality quantum numbers. The total charge operators corresponding to the fermionic doublet of the combined system are given by

\be
\widehat{\cal Q} (\beta )\,=\,{\cal Q} (\beta )\,-\,\widetilde{\cal Q} (\beta )\,,
\ee

\be
\widehat{\cal Q}^5 (\beta )\,=\,{\cal Q}^5 (\beta )\,-\,\widetilde{\cal Q}^5 (\beta )\,.
\ee

\no For the fermionic charge of the combined system we obtain the following selection rules 

\be
\big [ \widehat{\cal Q} (\beta )\,,\,\prod_{i = 1}^n \psi (x_i ; \beta) \prod_{j = 1}^{\tilde n} \widetilde\psi(y_j ; \beta ) \big ]\,=\,-\,(n\,-\,\tilde n)\, \prod_{i = 1}^n \psi (x_i ; \beta) \prod_{j = 1}^{\tilde n} \widetilde\psi(y_j ; \beta )\,,
\ee

\be
\big [ \widehat{\cal Q} (\beta )\,,\,\prod_{i = 1}^n \psi (x_i ; \beta) \prod_{j = 1}^{\tilde n} \widetilde\psi^\dagger (y_j ; \beta ) \big ]\,=\,-\,(n\,+\,\tilde n)\, \prod_{i = 1}^n \psi (x_i ; \beta) \prod_{j = 1}^{\tilde n} \widetilde\psi^\dagger (y_j ; \beta )\,.
\ee

\no This implies that in the massless free fermionic theory one finds the following off-diagonal $2n$-point functions 

\be
\langle 0 , \widetilde 0 \vert \prod_{i = 1}^n \psi (x^\pm_i ; \beta ) \prod_{j = 1}^n \widetilde \psi (y^\pm_j ; \beta ) \vert \widetilde 0 , 0 \rangle\,=\,\frac{\displaystyle \Big [ \prod_{i^\prime < i}^n (2 i \beta) \sinh \frac{\pi}{\beta}\big ( x^\pm_{i^\prime}\,-\,x^\pm_i \big ) \Big ]\, \Big [ \prod_{j^\prime < j}^n (2 i \beta) \sinh \frac{\pi}{\beta}\big ( y^\pm_{j^\prime}\,-\,y^\pm_j \big ) \Big ]}{\displaystyle
 \Big [ \prod_{i, j}^n (2 i \beta) \sinh \frac{\pi}{\beta}\big ( x^\pm_{i}\,-\,y^\pm_j\,-\,i\,\frac{\beta}{2}\,-\,i \epsilon \big ) \Big ]}
\,,
\ee

\no and

\be
\langle 0 , \widetilde 0 \vert \prod_{i = 1}^n \psi (x_i ; \beta ) \prod_{j = 1}^{n} \widetilde \psi^\dagger (y_j ; \beta ) \vert \widetilde 0 , 0 \rangle\,=\,0\,.
\ee

\no For the axial charge we obtain the selection rules (here $J = \psi_1^\dagger \psi_2 = \dt e^{2 i \sqrt \pi \phi}\dt\,,\,\widetilde J = \widetilde\psi_1^\dagger \widetilde\psi_2 = \dt e^{2 i \sqrt \pi \widetilde\phi}\dt $)

\be
\big [ \widehat{\cal Q}^5 (\beta )\,,\,\prod_{i = 1}^n J (x_i , \beta ) \prod_{j = 1}^{\tilde n} \widetilde J (y_j , \beta ) \big ]\,=\,- 2 (n\,-\,\tilde n )\,\prod_{i = 1}^n J (x_i , \beta ) \prod_{j = 1}^{\tilde n} \widetilde J (y_j , \beta )\,,
\ee

\be
\big [ \widehat{\cal Q}^5 (\beta )\,,\,\prod_{i = 1}^n J (x_i , \beta ) \prod_{j = 1}^{\tilde n} \widetilde J^\dagger (y_j , \beta ) \big ]\,=\,- 2 (n\,+\,\tilde n )\,\prod_{i = 1}^n J (x_i , \beta ) \prod_{j = 1}^{\tilde n} \widetilde J^\dagger (y_j , \beta )\,.
\ee

\no This provides  a  clear understanding of the physical meaning of the selection rule (\ref{SR}), that is, the only non zero contributions in the expansion (\ref{num4}) are those with zero total chirality  $\widehat{\cal Q}^5 (\beta )$

\be
\Big (\, \underbrace{N\,-\,N^\prime\,+\,\sum_{\ell = 1}^m \lambda_\ell}_{{\cal Q}^5} \,\Big )\,-\,\Big (\, \underbrace{\sum_{k = 1}^{\widetilde m} \tilde\lambda_k}_{\widetilde{\cal Q}^5}\, \Big )\,=\,0
\ee

\no of the combined system.

\section{Concluding Remarks}
\setcounter{equation}{0}

The main objective of this paper was to prove in thermofield dynamics the equivalence of the theory of massive free Fermi fields to the the sine-Gordon theory for a particular value of the sine-Gordon parameter $\kappa^2 = 4\pi$. Approaches of other authors \cite{Delep, Mintchev} differ from ours in two respects: i) they make use of the imaginary time formalism, and ii) treat the fermionic side in a hybrid way. We have treated the fermionic side of the problem strictly from the fermionic point of view. On the bosonic side we have used the thermofield bosonization \cite{ABR} in order to compute the n-point functions of the thermal chiral densities. On the bosonic side we first obtained the n-point functions of the chiral densities from a generalized Mandelstam operator ( with non-canonical scale dimension ) for the corresponding bosonized expressions at finite temperature, and then recovered from there the corresponding n-point functions of
the free theory as a limiting case. A gratifying byproduct of our analysis was the observation, that these n-point functions showed in a natural way, that the tilde fields
of the thermofield dynamics could be regarded as living on the lower branch 
of the integration contour in the complex time plane, displaced from the real time axis by $-\,i\frac{\beta}{2}$, in accordance with the work of ref.\cite{Mats1}.

We recognize that the ``revised version'' of thermofield dynamics formulation for fermions introduced in Ref. \cite{O}  enables to obtain the correct bosonized expression for the field $\widetilde\psi$ and  becomes crucial in order to establish the two-dimensional Fermion-Boson mapping within the thermofield dynamics approach, as well as, to obtain a clear understanding of the chiral selection rule of the combined system.

{\bf Acknowledgments}: We are grateful to Brazilian Research Council (CNPq) for partial financial support and to the FAPERJ(E-26/170.949/2005)-DAAD scientific exchange program which make this collaboration possible.

\newpage

\appendix{{\centerline{\bf{Appendix }}}}
\vspace{0.5cm}

\centerline{\bf Thermofield Bosonization of the Free Massless Fermion Field Revised}

\renewcommand{\theequation}{{A}.\arabic{equation}}\setcounter{equation}{0}

In this Appendix we shall consider the revised thermofield dynamics approach for fermions presented in Ref. \cite{O} in order to correct a mistake in our previous paper \cite{ABR} and give the revised thermofield bosonization prescription for the free massless Fermi field doublet $(\psi , - i {\widetilde\psi}^\dagger )$. Although in the fermionic formulation the use of the new version does not change the computation of the diagonal two-point functions of the free massless fields $\langle 0 , \widetilde 0 \vert \psi (x^\pm ; \beta ) \psi^\dagger (y^\pm ; \beta ) \vert \widetilde 0 , 0 \rangle$ and $\langle 0 , \widetilde 0 \vert \widetilde\psi (x^\pm ; \beta ) \widetilde \psi^\dagger (y^\pm ; \beta ) \vert \widetilde 0 , 0 \rangle$, it corrects a ``sign'' in our previous computation \cite{ABR} of the off-diagonal contribution $\langle 0 , \widetilde 0 \vert i \widetilde \psi (x^\pm ; \beta ) \psi (y^\pm ; \beta ) \vert \widetilde 0 , 0 \rangle$, and as we shall see, gives a new insight into the thermofield bosonization scheme.   

In the ``old version'', the operator taking one to the Fermion thermofields is given by \cite{TFD,U,Das,O}

\be\label{ouop}
U_F (\theta_F )\,=\,\exp \Big \{\,-\,\int_{-\infty}^\infty \,d\,p\,\theta_F (\vert p^1\vert  ,\beta )\Big ( \widetilde b (p^1)\,
b (p^1)\,-\,b^\dagger (p^1)\,\widetilde{b}^\dagger (p^1)\,+\, 
\widetilde d (p^1)\,
d (p)\,-\,d^\dagger (p^1)\,\widetilde{d}^\dagger (p^1)\,\Big ) \Big \}\,,
\ee

\no and the corresponding transformed annihilation operators are given by

\be\label{b1}
b (p ; \beta ) = b(p) \cos \theta_F (p ; \beta ) \,-\,\widetilde b^\dagger (p) \sin \theta_F (p ; \beta )\,,
\ee

\be\label{b2}
\widetilde b (p ; \beta ) = \widetilde b(p) \cos \theta_F (p ; \beta ) \,+\, b^\dagger (p) \sin \theta_F (p ; \beta )\,,
\ee

\no with similar expressions for $d$ and $\widetilde d$. As stressed in Ref. \cite{O}, the requirement for (\ref{b1}) and (\ref{b2}) to be consistent with each other implies the following ``tilde substitution rule'' for fermions  ($b \equiv b (p)$)

\be\label{tsr}
\widetilde{\widetilde b}\,=\,-\,b \,.
\ee

\no In the ``new version'' \cite{O}, the vacuum state $\vert 0 (\beta ) \rangle$ is obtained from the Fock vacuum $\vert \widetilde 0 , 0 \rangle$ by the  ``modified'' unitary operator ${\cal U}_F (\theta_F)$, which can be formally obtained from the old one (\ref{ouop}) by the substitution 

\be
\widetilde b \rightarrow i \widetilde b\,,
\ee
\be
\widetilde b^\dagger \rightarrow -\,i\,\widetilde b^\dagger\,,
\ee

\no with similar substitution for $\widetilde d$, i. e.,

\be
{\cal U}_F (\theta_F )\,=\,\exp \Big \{\,\mathbf{-\,i}\,\int_{-\infty}^\infty \,d\,p\,\theta_F (\vert p^1\vert  ,\beta )\Big ( \widetilde b (p^1)\,
b (p^1)\,{+}\,b^\dagger (p^1)\,\widetilde{b}^\dagger (p^1)\,+\, 
\widetilde d (p^1)\,
d (p)\, {+}\,d^\dagger (p^1)\,\widetilde{d}^\dagger (p^1)\,\Big ) \Big \}\,.
\ee

\no The revised fermionic transformations  are given by \cite{O}

\be
b (p ; \beta ) = b(p) \cos \theta_F (p ; \beta ) \,+\,i\,\widetilde b^\dagger (p) \sin \theta_F (p ; \beta )\,,
\ee

\be
\widetilde b (p ; \beta ) = \widetilde b(p) \cos \theta_F (p ; \beta ) \,-\,i\, b^\dagger (p) \sin \theta_F (p ; \beta )\,,
\ee

\no with similar expressions for $d$ and $\widetilde d$. This procedure replace the ``tilde substitution rule''  (\ref{tsr})  by the ``tilde conjugation operation''  

\be
\widetilde{(i b)} \,=\,-\,i\,\widetilde b \,,
\ee

\no analogous to the case of boson operators. Taking this into account, the revised expression for the two-dimensional free massless Fermi thermofields are given by 

$$
\psi (x^\pm ; \beta )\,=\,\frac{1}{\sqrt{2 \pi}}\,\int_0^\infty dp \,\Big \{\,f_p (x^\pm)\,\Big ( b (\mp p)\,\cos \theta_F ( p ; \beta )\,+\,i\,\widetilde b^\dagger (\mp p)\,\sin \theta_F ( p ; \beta )\,\Big )\,+
$$

\be\label{f1}
f^\ast_p (x^\pm)\,\Big ( d^\dagger (\mp p)\,\cos \theta_F ( p ; \beta )\,-\,i\,\widetilde d (\mp p)\,\sin \theta_F ( p ; \beta )\,\Big )\,\Big \}\,,
\ee

$$
\widetilde\psi (x^\pm ; \beta )\,=\,\frac{1}{\sqrt{2 \pi}}\,\int_0^\infty dp\, \Big \{\,f^\ast_p (x^\pm)\,\Big ( \widetilde b (\mp p)\,\cos \theta_F ( p ; \beta )\,-\,i\, b^\dagger (\mp p)\,\sin \theta_F ( p ; \beta )\,\Big )\,+
$$

\be\label{f2}
f_p (x^\pm)\,\Big ( \widetilde d^\dagger (\mp p)\,\cos \theta_F ( p ; \beta )\,+\,i\, d (\mp p)\,\sin \theta_F ( p ; \beta )\,\Big )\,\Big \}\,,
\ee

\no  where

\be
f_p (x)\,=\,e^{\,-\,i\,p x}\,.
\ee

Now, let us compute the off-diagonal two-point function within the fermionic version. Using (\ref{f1}) and (\ref{f2}), one gets \footnote{The Eq. (4.23) of Ref. \cite{ABR} is correct, but it cannot be obtained from (4.14) and (4.13) (where we have used the old version with (\ref{b1}) and (\ref{b2})), which gives 

$$
\frac{i}{\pi} \int_0^\infty \sin p (x^\pm - y^\pm ) N_F (\beta , p)\,e^{\,\frac{\beta p}{2}} d p\,,
$$

\no instead of $\cos p (x^\pm - y^\pm)$ in the integrand.}

\be\label{fc}
\langle 0 , \widetilde 0 \vert \,i\,\widetilde\psi ( x^\pm ; \beta )\,\psi (y^\pm ; \beta ) \vert \widetilde 0 , 0 \rangle\,=\,-\,\frac{1}{2 \pi}\,\int_0^\infty dp\,\frac{\cos p (x^\pm\, -\, y^\pm)}{\cosh \frac{\beta p}{2}}\,=\,-\,\frac{1}{2 \beta \cosh \frac{\pi}{\beta} (x^\pm\, -\, y^\pm )}\,,
\ee

\no which can be written as

\be\label{fc2}
\langle 0 , \widetilde 0 \vert \,i\,\widetilde\psi ( x^\pm ; \beta )\,\psi (y^\pm ; \beta ) \vert \widetilde 0 , 0 \rangle\,=\,-\,\frac{1}{2\,i\, \beta \sinh \frac{\pi}{\beta} (x^\pm\, -\, y^\pm \,-\,i\,\frac{\beta}{2})}\,.
\ee

\no In order to establish the Fermion-Boson mapping,  let us compute the two-point function above from the bosonized point of view. To this end, we define

\be
\psi (x^\pm ; \beta ) \,=\,\Big ( \frac{\mu}{2 \pi} \Big )^{\frac{1}{2}}\,e^{\,-\,\z (\mu^\prime ; \beta )}\,\dt\,e^{\textstyle\,2 \,i \,\sqrt \pi\,\phi (x^\pm ; \beta )}\,\dt\,,
\ee

\be
\widetilde \psi (x^\pm ; \beta ) \,=\,( - i )\,\Big ( \frac{\mu}{2 \pi} \Big )^{\frac{1}{2}}\,e^{\,-\,\z (\mu^\prime ; \beta )}\,K_\beta\,\dt\,e^{\textstyle\,2\, i\,\gamma\, \sqrt \pi\,\widetilde\phi (x^\pm ; \beta )}\,\dt\,,
\ee

\no where $K_\beta$ is a Klein factor \cite{Klein} which ensures normal anticommutativity between tilde fermion thermofield components and non-tilde ones \cite{ABR}, and the value of $\gamma = \,\pm\,1$ will be fixed at the end of the calculation. Using that \cite{ABR}

\be
\langle 0 , \widetilde 0 \vert \phi (x^\pm ; \beta ) \widetilde \phi (y^\pm ; \beta ) \vert \widetilde 0 , 0 \rangle\,=\,-\,\frac{1}{4\pi}\,f (\mu^{\prime\prime} , \beta )\,+\,\frac{1}{4 \pi}\,\ln \Big \{\,\cosh \big [ \frac{\pi}{\beta} ( x^\pm\, -\,y^\pm ) \big ] \Big \}\,,
\ee

\no we obtain (the global minus sign arises from the Klein factor)

\be
\langle 0 , \widetilde 0 \vert \,i\,\widetilde\psi (x^\pm ; \beta) \psi (y^\pm ; \beta ) \vert \widetilde 0 , 0 \rangle\,=\,-\,\frac{\mu}{2 \pi}\,e^{-\,2\,\z (\mu^\prime , \beta )\,+\,\gamma f (\mu^{\prime\prime} ,\beta )}\,\Big [\,\cosh \frac{\pi}{\beta} (x^\pm \,-\, y^\pm) \Big ]^{\,-\,\gamma}\,.
\ee

\no Using the asymptotic behavior

\be
\z (\mu^\prime \approx 0 , \beta ) \rightarrow\,\frac{1}{\beta \mu^\prime}\,+\,\frac{1}{2}\,\ln (\beta \mu^\prime)\,,
\ee

\be
f (\mu^{\prime\prime} \approx 0 ,  \beta ) \rightarrow \frac{2}{\beta \mu^{\prime\prime}}\,,
\ee

\no we get

\be
\langle 0 , \widetilde 0 \vert \,i\,\widetilde\psi (x ; \beta) \psi (y ; \beta ) \vert \widetilde 0 , 0 \rangle\,=\,-\,
\Big (\frac{\mu}{ \pi \mu^\prime} \Big )\,e^{\,\frac{2}{\beta} \big ( \frac{\gamma}{\mu^{\prime\prime}}\,-\,\frac{1}{\mu^\prime}\big )}\,\frac{1}{2 \beta\,\big [\,\cosh \frac{\pi}{\beta} (x^\pm\, -\, y^\pm) \big ]^{\,\gamma}}\,.
\ee

\no In order to recover the same space-time dependence as in (\ref{fc}) we must require that 

\be
\gamma\,=\, 1\,.
\ee

\no In order to obtain an infrared cut-off independent two-point function,  the free massless scalar thermofield theory should be considered as the zero mass limit of the massive free scalar thermofield 

\be
\langle 0 , \widetilde 0 \vert \Sigma (x ; \beta ) \Sigma (y ; \beta ) \vert \widetilde 0 , 0 \rangle_{m \rightarrow 0}\,\rightarrow
\langle 0 , \widetilde 0 \vert \phi (x ; \beta ) \phi (y ; \beta ) \vert \widetilde 0 , 0 \rangle\,,
\ee

\be
\langle 0 , \widetilde 0 \vert \widetilde\Sigma (x ; \beta ) \widetilde\Sigma (y ; \beta ) \vert \widetilde 0 , 0 \rangle_{m \rightarrow 0}\,\rightarrow
\langle 0 , \widetilde 0 \vert \widetilde\phi (x ; \beta ) \widetilde\phi (y ; \beta ) \vert \widetilde 0 , 0 \rangle\,,
\ee

\be
\langle 0 , \widetilde 0 \vert \Sigma (x ; \beta ) \widetilde\Sigma (y ; \beta ) \vert \widetilde 0 , 0 \rangle_{m \rightarrow 0}\,\rightarrow
\langle 0 , \widetilde 0 \vert \phi (x ; \beta ) \widetilde\phi (y ; \beta ) \vert \widetilde 0 , 0 \rangle\,.
\ee

\no In this way, the infrared regulator $\mu$ of the zero temperature two-point function should be identified with the infrared cut-offs $\mu^\prime$ and $\mu^{\prime\prime}$ of the temperature-dependent contributions $\z (\mu^\prime , \beta )$ and $f (\mu^{\prime\prime}, \beta )$, i. e.,  

\be
\mu^{\prime\prime} = \mu^\prime = \frac{\mu}{\pi}\,,
\ee 

\no and we get,

\be
\langle 0 , \widetilde 0 \vert \,i\,\widetilde\psi (x ; \beta) \psi (y ; \beta ) \vert \widetilde 0 , 0 \rangle\,=\,-\,
\frac{1}{2 \beta\,\cosh \frac{\pi}{\beta} (x - y) }\,,
\ee

\no in accordance with (\ref{fc}). Since the off-diagonal selection rule carried by the Wick-ordered exponential requires that $\gamma = 1$, one concludes that the bosonized expression for $\widetilde\psi (x)$ is not obtained just by the tilde conjugation operation ($\widetilde W$) of the corresponding Wick-ordered exponential $W$ defining $\psi (x)$. Besides a multiplicative factor $(- i)$ and  Klein factors, the bosonized version of the field $\widetilde\psi (x)$ is obtained only by the tilde conjugation of the creation and annihilation components of the field $\phi (x)$ at the exponent, which can be achieved by defining $\widetilde \psi$ in terms of the Wick exponential $\widetilde{W^\ast}$,

\be
i\,\widetilde\psi (x ; \beta )\,=\,U_B (\theta_B)\,i\,\widetilde\psi (x) U_B^{\,-\,1} (\theta_B)\,,
\ee

\no with

\be\label{psitildebos}
i\,\widetilde \psi (x) \,\doteq\,\Big ( \frac{\mu^D}{2 \pi} \Big )^{\frac{1}{2}}\,K\,\widetilde{W^\ast} (x)\,=\,
\Big ( \frac{\mu^D}{2 \pi} \Big )^{\frac{1}{2}}\,K\,
\dt\,e^{\textstyle\, i \,2 \,\sqrt \pi\,\widetilde\phi (x)}\,\dt\,.
\ee

One concludes that the revised version of the thermofield dynamics formulation for fermions introduced in Ref. \cite{O} plays an important role in order to establish the two-dimensional Fermion-Boson mapping within the thermofield dynamics approach. As a byproduct, according with Eq. (\ref{fc2}), this formulation enable to regard the tilde fields 
of the thermofield dynamics as living on the lower branch 
of the integration contour in the complex time plane, displaced from the real time axis by $-\,i\frac{\beta}{2}$, in accordance with the work of ref.\cite{Mats1}.

\end{document}